*Research Article*

# A Genetic Algorithm-Based Support Vector Machine Approach for Intelligent Usability Assessment of m-Learning Applications

**Muhammad Asghar, Imran Sarwar Bajwa, Shabana Ramzan, Hina Afreen, and Saima Abdullah**

*Department of Computer Science, Islamia University of Bahawalpur, Bahawalpur, Pakistan*

Correspondence should be addressed to Imran Sarwar Bajwa; imran.sarwar@iub.edu.pk

In the field of human computer interaction (HCI), the usability assessment of m-learning (mobile-learning) applications is a real challenge. Such assessment typically involves extraction of best features of an application like efficiency, effectiveness, learnability, cognition, memorability, etc., and further ranking of those features for overall assessment of the quality of the mobile application. In the previous literature, it is found that there is neither any theory nor any tool available to measure or assess a user's perception and assessment of usability features of a m-learning application for the sake of ranking of the graphical user interface of a mobile application in terms of a user's acceptance and satisfaction. In this paper, a novel approach is presented by performing a mobile application's quantitative and qualitative analysis. Based on the user's requirements and perception, a criterion is defined based on a set of important features. Afterwards, for the qualitative analysis, genetic algorithm (GA) is used to score prescribed features for usability assessment of a mobile application. The used approach assigns a score to each usability feature according to the user's requirement and weight of each feature. GA performs the rank assessment process initially by performing feature selection and scoring the best features of the application. A comparison of assessment analysis of GA and various machine learning models, i.e., K-nearest neigbors, Naïve Bayes, and Random Forests is performed. It was found that GA-based support vector machine (SVM) provides more accuracy in the extraction of best features of a mobile application and further ranking of those features.

## 1. Introduction

The e-learning applications are emerging as an alternate technology to the classroom leaning specifically in the last couple of years where COVID-19 has affected almost every field of life including education. However, the users of mobile e-learning applications face a problem: its usability in terms of providing user-friendly mobile learning applications to provide an optimum mobile learning experience. In HCI, five main components can define usability listed in research [1]: learning, efficiency, memorability, general accuracy, and user satisfaction. The usability features typically focus on the concept of ease of use as mentioned in the literature. In an easy-to-use application, a user can do tasks easily. Hence, the focus is to improve user task efficiency and minimizing complexity in the user interface. The basic HCI principles suggest improvement of user satisfaction by making mobile learning applications engaging, attractive and aesthetically pleasing. To check the efficiency of mobile e-learning systems, the evaluation of e-learning systems can be done by assessing its usability.

Distance education has unique benefits. It provides a winning strategy to address specific needs such as overcrowded education facilities. It may also support the students and teachers anytime and anywhere specifically who live far from schools or universities. The learning material can be share via e-learning applications and it can be a valuable resource for students, specifically the students with disabilities. E-learning is the latest method of remote learning by distributing online learning materials and processes [2]. Detailed data and services are provided to the users, such as cultural circumstances, technological experiences, facilities, and physical/cognitive skills. It is essential to provide e-learning facilities to learners all over the country to minimize digital divide and gap, socially and culturally [3]. Ensuring compatibility and access to the maximum number of users should is a key objective for developers of e-learning applications with a requirement to make it possible for all users to use such applications easily and effectively. A few factors contribute to the usability of a mobile app that impacts the overall efficiency and efficiency with which a user achieves ones' objectives [4]. A useable interface should have three main features as follows:

(i) The easy familiarity with the user interface (UI) and its working

(ii) Users may achieve their goal easily by using the application

(iii) The applications must be error-free

Usability problems become more critical by user interface errors, in which every step or click requires troubleshooting rather than smooth running of the application. Mobile users also may experience compatibility problems when installing third-party tools, since the system may not support various formats and data transfer methods may be complicated. Users may also face usability issues due to a lack of responsive interface when interacting with the multiple devices. Moreover, while logging into a mobile learning application, a new varying interface may become complex for a user [5].

Various problems identified in the research [5–8] are related to the quality of usability features that impact the functionality and usability of that application. This study further investigates the problem of less learnability of students due to difficulty in using such e-learning applications. In this study, a quantitative analysis is performed to collect a user's (a student's) perspective of usability requirements and perception of quality of a m-learning application's user interface. The proposed approach studies the problem of low usability by applying conventional usability evaluation strategies along GA-based SVM and other generative machine learning models [6]. This study helps identify and prioritize key problems and issues related to the usability features for m-learning applications and helps find a holistic approach of assessing usability issues with a mobile learning application. In this paper, a novel approach is presented that uses genetic algorithm (GA) based support vector machine (SVM) to extract prominent usability features of a mobile application as per user's requirement and then scores these features to rate quality of a user interface of a mobile application.

The rest of the paper is structured, as follows: Section 2 discusses the literature discussing usability issues and challenges and their possible solutions; Section 3 describes the used approach and the genetic algorithm-based support vector machine algorithm for prediction of usability issues; Section 4 provides details of the experiments, their results and discussions on the working and performance of the proposed approach, and Section 5 concludes the presented approach and achieved results.

## 2. Literature Review

In this section, the previous research and approaches available in the literature are discussed, focusing on usability issues and critical review to identify the research gap. Mobile e-learning applications were initially introduced in 2014 in Pakistan for primary classes such as 1 to 3 grades. A manual system was used for evaluating the performance of students before the start of usage of mobile learning applications [1, 9–11]. Before using mobile learning applications, the tests of students were conducted manually on paper and were checked manually and it was a complex and time-consuming process. However, with the introduction of the mobile e-learning applications helped in solving various problems related to notes' management, student's evaluation, grading assessment uploading, policies, and the services of mobile learning applications that were graded to $8^{th}$ grade students with a few new features in near future [8].

A learning management system (LMS) is useful in online providing educational materials, management of learning process for the sake of convenience of the users including students, teachers, learners, and content creators. To check the efficiency of e-learning systems, the evaluation of e-learning systems can be done by assessing its usability. It can be assessed that how well tools and technologies are working for users. A major part of e-learning systems is the learning management system (LMS). It can be beneficial when evaluated instead of satisfaction of users and ease of use by its functionality [12]. The dream of m-education is made more practical by the availability of affordable devices and low-cost Internet packages with the launch of 3G and 4G technologies in Pakistan [13]. In Pakistan, the school systems are almost physical first time due to COVID-19.

Additionally, systems like Literacy and Numeracy Drives (LND) in 2016 were being used in public institutes for grade III that the government of Punjab introduced. This application contained Multiple Choice Questions (MCQs) of various subjects like Mathematics, Urdu, and English [11, 14]. Additionally, IUB-LMS (LMS-IUB) was introduced in 2019 and is successful system in terms of functionality. This e-Learning system was introduced through a mobile Application named as "LMS-IUB PUBLIC" version 1.1.1.

To improve the quality of education in Pakistan, various fundamental actions are being taken to ensure the check and balance of education quality and the achievement of goals to facilitate the students in learning various skills. In 2021, a new version of LMS-IUB is introduced as LMS-IUB version 1 (application) that is successfully implemented in IUB, Bahawalpur Pakistan and achieved its educational objectives featured at IUB website. Reference [12, 13]. In Pakistan, the major population belongs to the rural and underdeveloped areas. The people living in rural areas cannot afford or use digital devices due to lack of technology, infrastructure, or financial issues. That is why Pakistan faces multiple challenges in implementing a quality e-learning education system in Pakistan [2, 14, 15]. In Table 1, a list of adaptive features is shown that describes significance of studying assessment mechanisms of usability in our study. Table 1 also highlights the proportion of each usability feature with respect to its significance and discussion in the referenced research and previous work.

Table 1: List of selected features for usability concerning its significance.

| Attribute | Share | References |
|---|---|---|
| Efficiency | 37 (70%) | Almarashdeh et al.; Farmanesh et al.; Aydin et al. |
| Satisfaction | 35 (66%) | Wagner et al.; Junus et al. |
| Effectiveness | 31 (58%) | Kipkurui et al., "Evaluating usability of E-Learning systems in universities" |
| Learnability | 24 (45%) | Kipkurui et al., "Evaluating usability of E-Learning systems in universities" |
| Memorability | 12 (23%) | Kumar et al.; Inversini et al. |
| Errors | 9 (17%) | Zaharias |
| Simplicity | 7 (13%) | Hsieh and Koong Lin |
| Ease of use | 7 (9%) | Conley et al. |

It is also studied in the previous literature that mobile phones can distract students, for example, by calling, checking social media updates such as Facebook, twitter, etc. It is why some countries have prohibited mobile telephones from entering during class time to avoid any possible disturbance [16, 17]. For decades, technology has taken various steps towards quality education. The mobile revolution is fundamentally changing other fields of life, education is also being affected with a new tool that supplies knowledge. M-education typically involves the learning process through personal electronic devices and gadgets such as smartphones, tablets, and social media. Mobile learning is the most creative and the easiest way to learn in today's world. The percentage of studies for the significance of usability of mobile apps is shown in Table 2.

The traditional education system in Pakistan is flawed and major reforms are required such as usage of modern technologies and electronic devices. Concerning the Education Development Index (LDI), Pakistan need to improve its education quality with the help of modern technologies and online platform. Table 3 compares various techniques and methods used by authors such as black/white board, m-Learning, ELMS and LMS. Here, a few features of these techniques are compared to find out a better one used in the previous research. . On the base of statistics shown in Table 3, it is found that there is a need of a better design and heuristics-based e-learning application that the larger audience can easily adapt.

It is shown in Table 3 that various e-learning applications specifically majorly lack in interaction, memorability, and consistency while these applications minorly lack in efficiency and ease of use. After identifying the research gap, it was also found that there is need of a theoretical approach that can assess quality of various usability features such as interaction, efficiency, memorability, consistency, ease of use, etc. Such assessment will help GUI designers to design better and improved interface of e-learning applications for the enhanced experience of the users. After considering the research gap as shown in Table 3, Genetic algorithm (GA) based support vector machine (SVM) are used to extract prominent usability features as per user's need. These features are scored to rate user-interface quality of a mobile application. The study will focus the following objectives:

(i) To identify the usability issues in a m-learning application

(ii) To propose a new and improved model of usability assessment of a m-learning application

(iii) To develop a prototype tool with the purposed model as a proof of concept

(iv) To evaluate the proposed model for its effectiveness and correctness

A specific focus of our study is learnability related issues with e-learning applications. Therefore, there is a need to perform a usability evaluation of selected mobile learning applications to find out the quality of human interaction of these mobile e-learning applications. There is also a need to analyse how a poor user interface can affect learning of a student and how an improved user interface can help students in better learning with ease and effectivity. Table 3 shows the comparison of previous studies and highlights the research gap.

In the literature it is also identified that, the technical features (as shown in Table 3) majorly contribute in a substantial and durable building block and provides versatility in functionality. Additionally, mobile networking and technical elements can be added in an application to enhance smart phone functionality. The use of mobile phones in schooling can be implemented with the idea of m-learning technology. The typical m-Learning technology provides a modern learning platform where content can be quickly obtained through a smartphone and interaction between a student and a teacher becomes easy and affordable. However, assessing the usability issues of the current m-learning application in terms of user satisfaction is still an open challenge. The user's adaptive standards assess its usability, comfort in screen reading, functionality achievement, user satisfaction and learning capacity. The factors mentioned above affect the availability, adaptation, and page presentation of a website structure's ease of use and navigation. Initially, an open curriculum was launched in 2002 by the Massachusetts Institute of Technology (MIT) with an idea of e-learning [14]. A few images or textures on the website were used that had no context to make users feel cognitively well. Moreover, creating an evaluation e-learning model aims to reduce the gap between the conceptual model of the user and the creator's experience. The relationship between user and device is the user's conceptual model [14, 17, 19–21].

Table 2: Percentage of studies concerning usability attributes of typical m-learning applications.

| Attribute | Studies before 2020 | Studies in 2020-21 | References |
|---|---|---|---|
| Effectiveness | 58 | 60 | Almarashdeh et al.; Farmanesh et al.; Aydin et al. |
| Efficiency | 70 | 75 | Wagner et al.; Junus et al. |
| Satisfaction | 66 | 78 | Kipkurui et al., “Evaluating usability of E-Learning systems in universities” |
| Error | 65 | 78 | Kipkurui et al., “Evaluating usability of E-Learning systems in universities” |
| Cognition | 46 | 76 | Kumar et al.; Inversini et al. |
| Learnability | 76 | 90 | Zaharias |

Cognitive science has typically influenced the science of human computer interface to better user experience. There are a lot of previous studies that are focusing on modelling a mental process. However, epistemological and methodological issues have remained untouched and are not considered while developing a user interface. Scientifically, a cognitive load is a mixture of inherent load, and external loads [22]. Cognition is the subject of modification. The idea of extended cognition will undergo yet another transition, which relates to the way people use technical and cognitive artefacts to enhance their ability to perform a functionality with ease. One of the key disadvantages of lack of usability assessment mechanism is difficulty measuring the difference in a user's mental model from the designer's perceptions. Buchner et al. showed that experimental methodology uses robots and essential user interface [14]. The user's experience is an important problem and leads to cognitive stress [12]. Asarbakhsh and Sandar are the two key contributors in enhancing education and learning utilizing technology in medical education using VLEs [15]. They estimated that an e-learning system would classify 85 per cent of the problems through a questionnaire/survey and test techniques [23]. The user experience is progressively expanding and changing. Disability accessibility problems impacting educational users. Freire et al. help technologies are influenced by physically disabled people using websites [16].

Hasan implemented guidelines for assessing the usefulness of online learning management systems, combined with an easy navigation framework [22]. The students used a questionnaire for Jordan University and two automated online tools, the html Toolbox and a web page analyser. Each part covers the usefulness of work presented by Aslam et al. [18]. Additionally, Jooet et. al. [17] had developed a model for adaptive learning, and usability of educational content management systems [21]. Their research aims to develop usability evaluation models and a survey tool to measure academic library websites' capacity, usability, and efficiency. Moreover, various scenarios, tasks, individual and collaborative tasks and actions were key components of the task model. Launching new technology in the market presents various risks, including more time consumption, task failure, and usage of extra effort [24].

## 3. Used Approach

The approach used in current research work is based on a simple model. The proposed approach is used for defining the relationship among a system, a user and a designer. The model developed in this research is based on an improved approach used to assess an m-learning application. In the experiments, a case study of LMS-IUB model was used to evaluate the performance of the proposed approach. The proposed approach is based on previously developed models of cognition. The gap between the mental model of users and the perception of designers can be reduced by conducting surveys, interviews and feedback from teachers and students in terms of their perception with a m-learning application. Figure 1 shows the proposed model for intelligent assessment of usability of m-learning applications. Survey, feedback and interviews are the tools used to fill the gap between designer and user. Application user satisfaction depends on many factors, and usability is one of them. A website should be logically organized to help the user achieve his goal.

This paper proposes an automated and intelligent method for usability testing of a m-learning application using machine-learning techniques to uncover usability issues and problems faced by the users while interacting with that particular m-learning application. The architecture of the used approach is shown in Figure 1 that takes input of the users of a m-learning application to find users' particular needs and requirements. The proposed approach genetic algorithm-based support vector machine algorithm to extract key usability features and rank the features as per usability quality. For further evaluation, the proposed approach is also experimented with random forest, decision trees, models of regression, etc. A range of machine learning patterns are tested with 10-fold cross validation on different data sets to assess the best possible model for a particular application's assessment. Finally, a recommendation is also generated to improve the usability of the application by generating a list of influential features and characteristics.

Figure 1 shows the workflow of the proposed approach. First, the user's point of view regarding user's needs and requirements are collected with the help of a questionnaire-based survey. A few datasets based on the observations and tests of different users are used to collect human points of view. Similarly, an ontology having an expert's opinion is used to rate the quality if identified features of a m-learning application. In the used approach, a machine learning-based model is used for clustering of dominant features. For this purpose, experimentation is done with GA-based support vector machine, Naïve Bayes, KNN Algorithm, and Random Forests. For the feature selection in usability testing, the usability data of a m-learning application is compared with a HCI ontology to score the usability features. A list of features is extracted after data analysis by using expert's opinion observations. The details of each step of the used approach are given as follows:

Table 3: Comparison of different previous techniques in terms of technical and usability features of applications (https://ieeexplore.ieee.org/document/9042272).

| | | Technical features | | | | | | Usability features | | | | |
|---|---|---|---|---|---|---|---|---|---|---|---|---|
| Ref | Application | Battery life | Resolution | Portability | Connectivity | Bandwidth | Screen and key size | Interaction | Efficiency | Memorability | Consistency | Ease of use |
| [1] | Blackboard | ✓ | ✓ | ✓ | ✓ | High | Adjustable | ✓ | ✓ | ✗ | ✗ | ✓ |
| [4] | M learning | ✓ | ✓ | ✓ | ✗ | High | Nonadjustable | ✓ | ✓ | ✗ | ✗ | ✓ |
| [3] | ELMS | ✗ | ✓ | ✓ | ✗ | High | Adjustable | ✓ | ✓ | ✓ | ✓ | ✓ |
| [9] | LMS | ✓ | ✗ | ✓ | ✗ | Low | Nonadjustable | ✗ | ✗ | ✗ | ✓ | ✓ |
| [15] | ELMS | ✓ | ✗ | ✓ | ✗ | Low | Nonadjustable | ✗ | ✓ | ✗ | ✓ | ✓ |
| [18] | Learning | ✗ | ✓ | ✗ | ✓ | Low | Nonadjustable | ✓ | ✓ | ✓ | ✗ | ✗ |
| [19] | Mobile internet LMS | ✗ | ✓ | ✗ | ✓ | Low | Adjustable | ✓ | ✗ | ✓ | ✗ | ✗ |
| [20] | ELMS | ✗ | ✓ | ✗ | ✓ | Low | Adjustable | ✓ | ✓ | ✗ | ✗ | ✗ |
| [17] | E-learning | ✓ | ✓ | ✗ | ✗ | Low | Adjustable | ✗ | ✓ | ✓ | ✗ | ✗ |

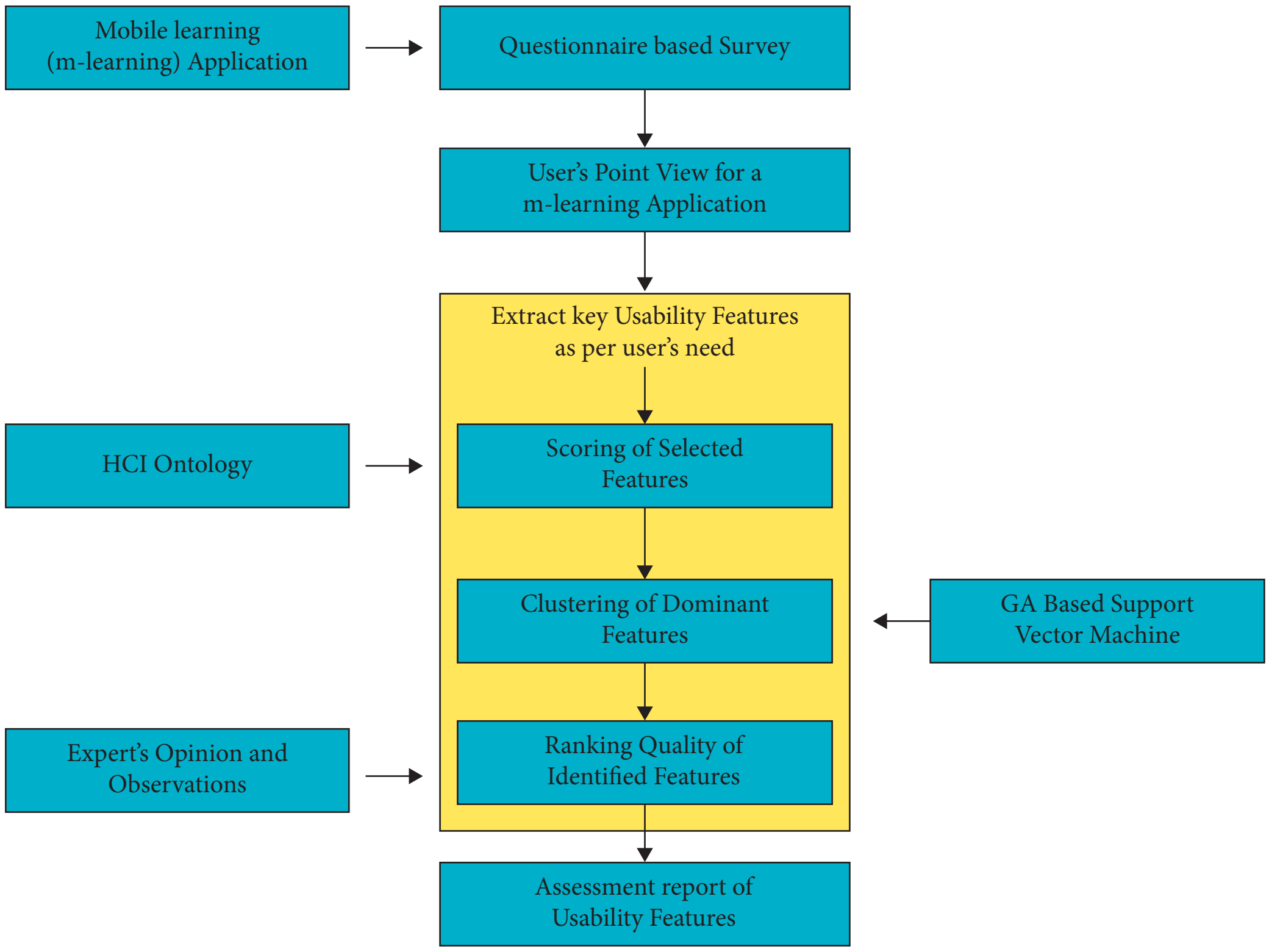

Figure 1: Proposed model for usability assessment of a m-learning application.

*3.1. Input Acquisition.* To find a user's point of view for a m-learning application, a questionnaire-based survey was performed for the usability assessment of the IUB-LMS system. In this study, the students of the Islamia University of Bahawalpur were involved as respondents.

There were 19 questions in the questionnaire addressing various usability issues that can be important for any user of a m-learning application. The responses of the survey were collected using an online Google form.

*3.1.1. Used m-Learning Application for the Study.* The E-learning system of IUB was launched in spring 2020. Figure 2 is the IUB-LMS system which shows a screenshot of the LMS of IUB. The focus of this study was to identify the dominant usability features of the LMS-IUB application and evaluate and score with the proposed approach for assessment of the usability features. Figure 2 shows the interface of LMS-IUB. The LMS-IUB is based on CSS and Laravel platform. This m-learning application provides a graphical user interface that helps students and teachers directly interact.

There are 40,000 students in IUB and all are registered at LMS-IUB application. A sample of 200 students was selected from different departments according to G-formula. The users were grouped into three types: primary, secondary, and tertiary users. Primary users are students, Secondary users are Teachers and tertiary users are other nonacademic staff.

*3.1.2. Population and Samples.* There are a total of 40000 students in the Islamia University of Bahawalpur. The dataset contains 200 students (samples of 500) of different departments of Islamia University of Bahawalpur. The excel sheet responses are used for data analysis. There are two types of questions, i.e. Open-ended (universal) and Closed-ended (individual) questions in the questionnaire. Figure 3 shows the proportion of different departments from which people responded:

Table 4 indicates the qualitative and quantitative analysis and the number of usability assessment methods (UEMs) applied to each usability attribute. The responses of the students is then coded in an excel sheet.

*3.2. Extracting Usability Features.* Extracting usability features from quantitative data is challenging because gathering respondents' information was not easy. The quantitative analysis of questionnaires provides the knowledge and capability for a greater understanding of choice of decisions. After asking questions from different students we have collected the usability information in descriptive and visual forms. Following Table 5 shows the number of polarity of each response from students:

We have taken the following two graphs from Google response forums to show students' response for the feature of ease of use and effectiveness of IUB LMS. Figure 4 shows the responses of students and their perceptions:

Figure 5 shows the response of students and the comparison of all features with all the data labels. The graph highlights that the majority of the respondents do care about efficiency and memorability. Afterwards, they give importance to learnability and ease of use along effectiveness of the system. Moreover, cognition and consistency are not highly demanded features for IUB-LMS system.

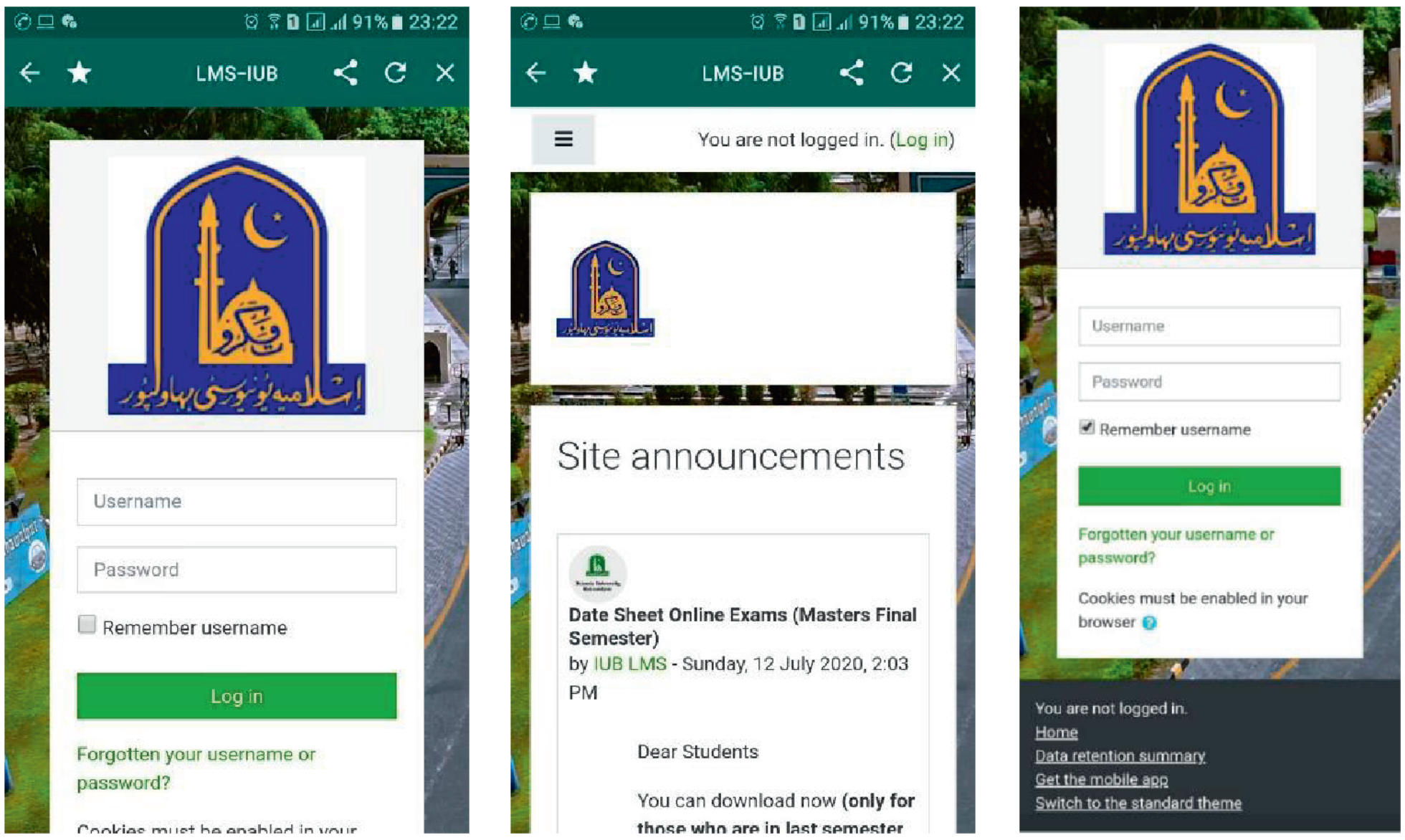

Figure 2: Login interface of LMS-IUB (https://m.apkpure.com/lms-iub-the-islamia-university-of-bahawalpur/com.iub.lms_iub).

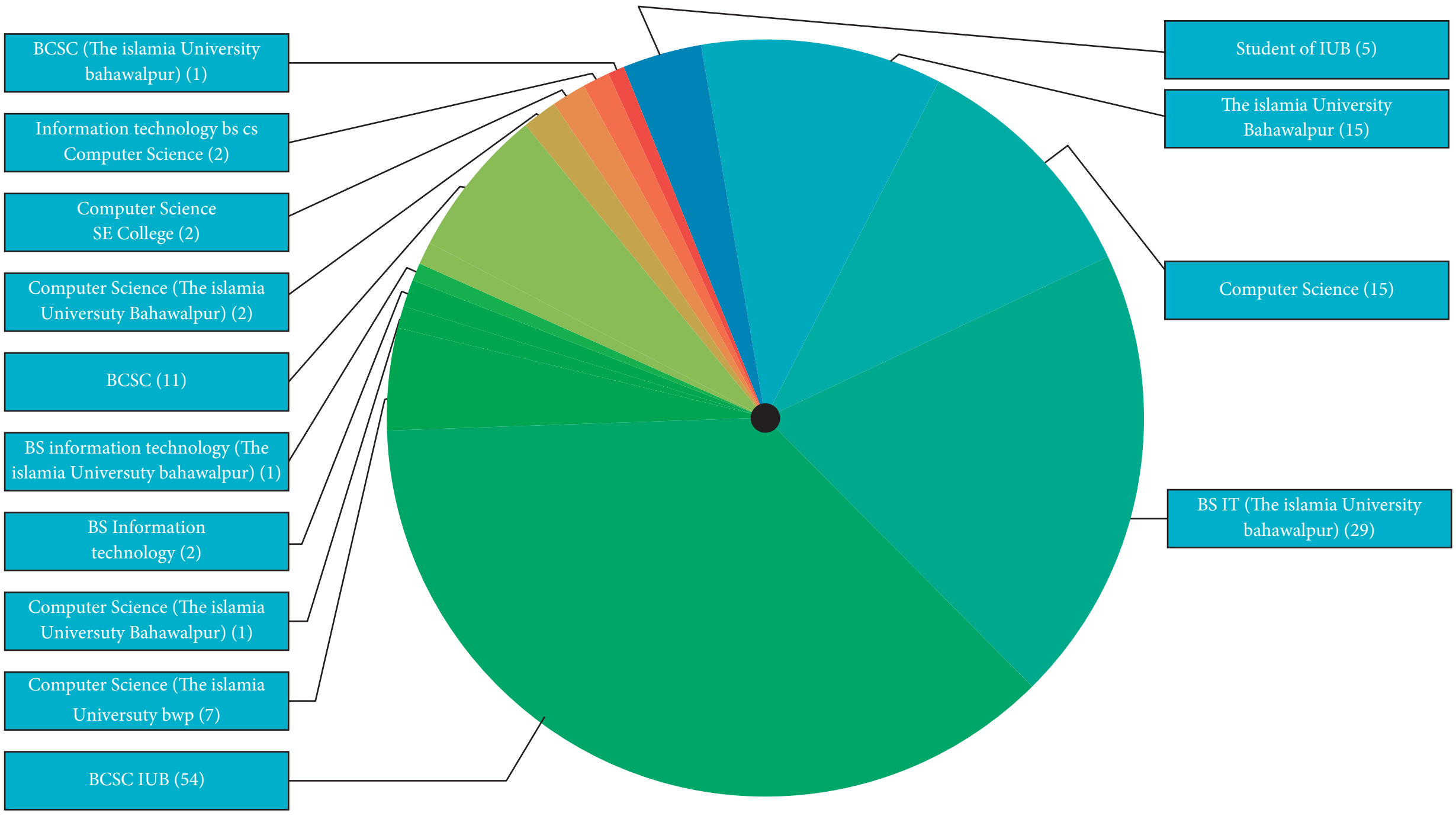

Figure 3: Graph explains the proportion of different departments from which people responded.

Table 4: The number of occurrences of usability evaluation methods (UEMs) applied to particular usability attributes.

| Attribute | Survey | Controlled observation | Eye tracking | Thinking aloud | Interview |
|---|---|---|---|---|---|
| Efficiency | 3 | 8 | 0 | 0 | 0 |
| Satisfaction | 1 | 0 | 0 | 0 | 0 |
| Effectiveness | 5 | 6 | 0 | 0 | 0 |
| Learnability | 3 | 5 | 0 | 0 | 0 |
| Memorability | 4 | 3 | 1 | 1 | 0 |
| Cognitive load | 0 | 1 | 1 | 1 | 0 |
| Errors | 1 | 4 | 9 | 0 | 0 |
| Simplicity | 4 | 2 | 9 | 0 | 1 |
| Ease of use | 2 | 0 | 9 | 0 | 0 |

Table 5: Response polarity.

| Features | 1: strongly agree | 2: agree | 3: neutral | 4: disagree | 5: strongly disagree |
|---|---|---|---|---|---|
| Efficiency | 48 | 22 | 21 | 8 | 7 |
| Effectiveness | 38 | 26 | 18 | 6 | 9 |
| Ease of use | 39 | 28 | 18 | 8 | 7 |
| Learnability | 41 | 27 | 21 | 4 | 5 |
| Memorability | 29 | 10 | 9 | 7 | 13 |
| Cognition | 12 | 9 | 7 | 5 | 13 |
| Consistency | 9 | 7 | 6 | 4 | 14 |

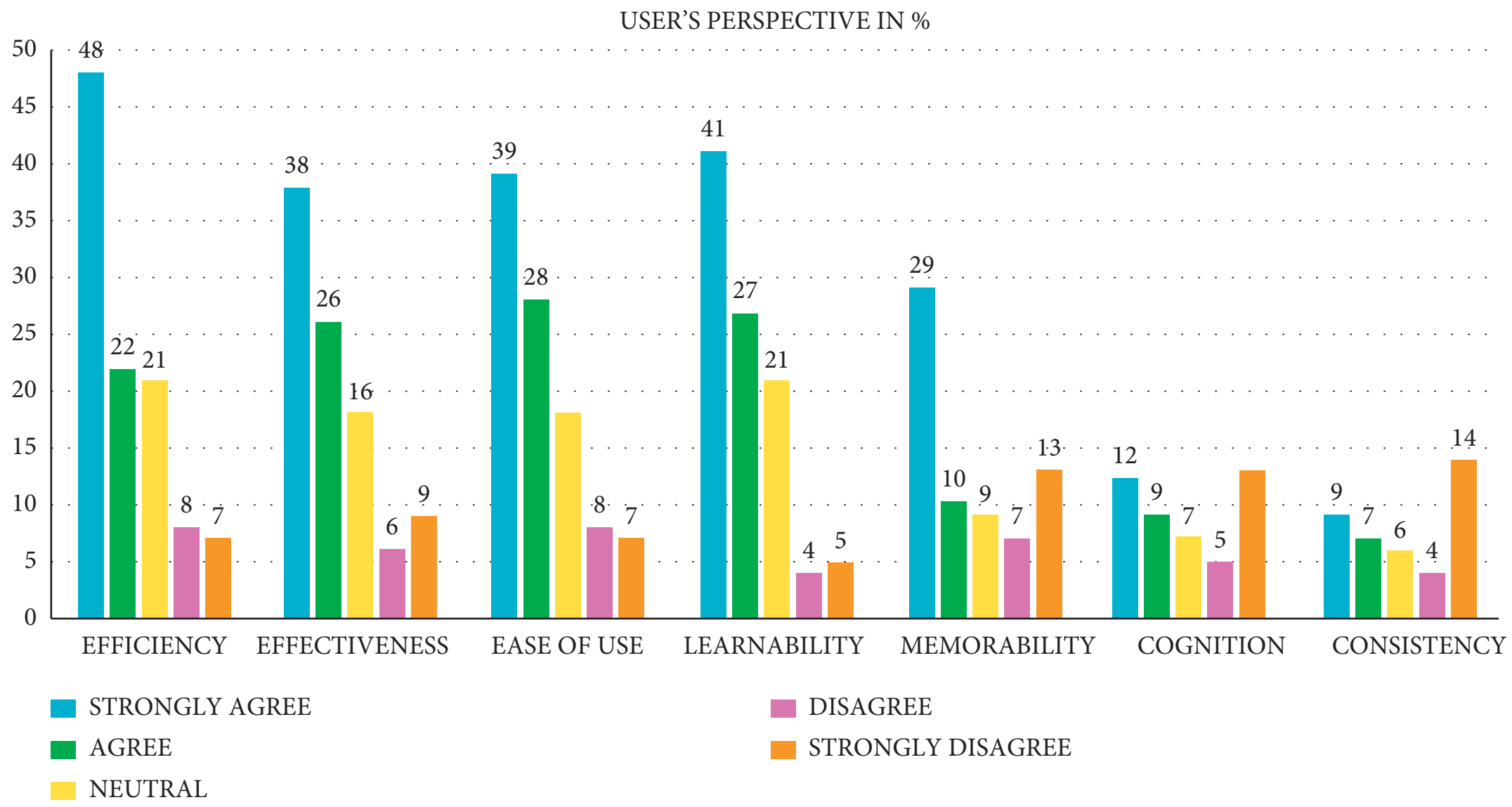

Figure 4: Response for the question in % concerning user's perspective.

Figure 6 shows the diversity in responses to each question of the survey. Here, it is shown that each question's detail concerning number of respondents is presented.

*3.3. Scoring of Features (GA-Based Scoring).* While it has many benefits, it can be a deep problem to scoring the functionality. Indeed, in many cases the scores provided by the various filters can be incomplete and similar features can be abruptly scored. Firstly, we regard the grade aggregation problem as an optimization problem, in which an optimal list can be found which approximates all aggregated lists.

In Figure 7, we explain the working flow of research method that how to figure out the main features and scoring of the features using genetic algorithm which are most important in usability.

We then concentrate on the problem of unjointed score, then perform a new algorithm, eliminating unjointed score for similar features, and removing features which give the target concept less details. Figure 6 shows our approach that output was evaluated using four credit data sets and 1 to 3 filters and four well-known aggregation techniques were checked. Scoring of features is the rank of features selected by the genetic algorithm. Following is the mathematical notations that how we have used GA for feature scoring.

*3.3.1. Initialization of Population.* We have selected attributes (ref. Table 2) and initialized as a population of genetic algorithm. The central difference of scoring between each feature in discretization form of two-dimensional, Matrix are defined as $U_{i\cdot j}$: where residual nodes are $i$ and $j$.

*3.3.2. Input Features*

$$ui, j = g(ui + 1, ui - 1, j, ui, j + 1, ui, j - 1). \quad (1)$$

*3.3.3. Node Weights.* The nodal residue for node $(i, j)$ is defined as

$$ri, j = ui, j - g(ui + 1, j, ui - 1, j, ui, j + 1, ui, j - 1). \quad (2)$$

*3.3.4. Weight of Each Node.* The overall residue for equation is the sum of squares of the nodal residues is stated as

$$R = \sqrt{nx} \sum i = 1\, my \sum j = 1\, r2i, j \quad (3)$$

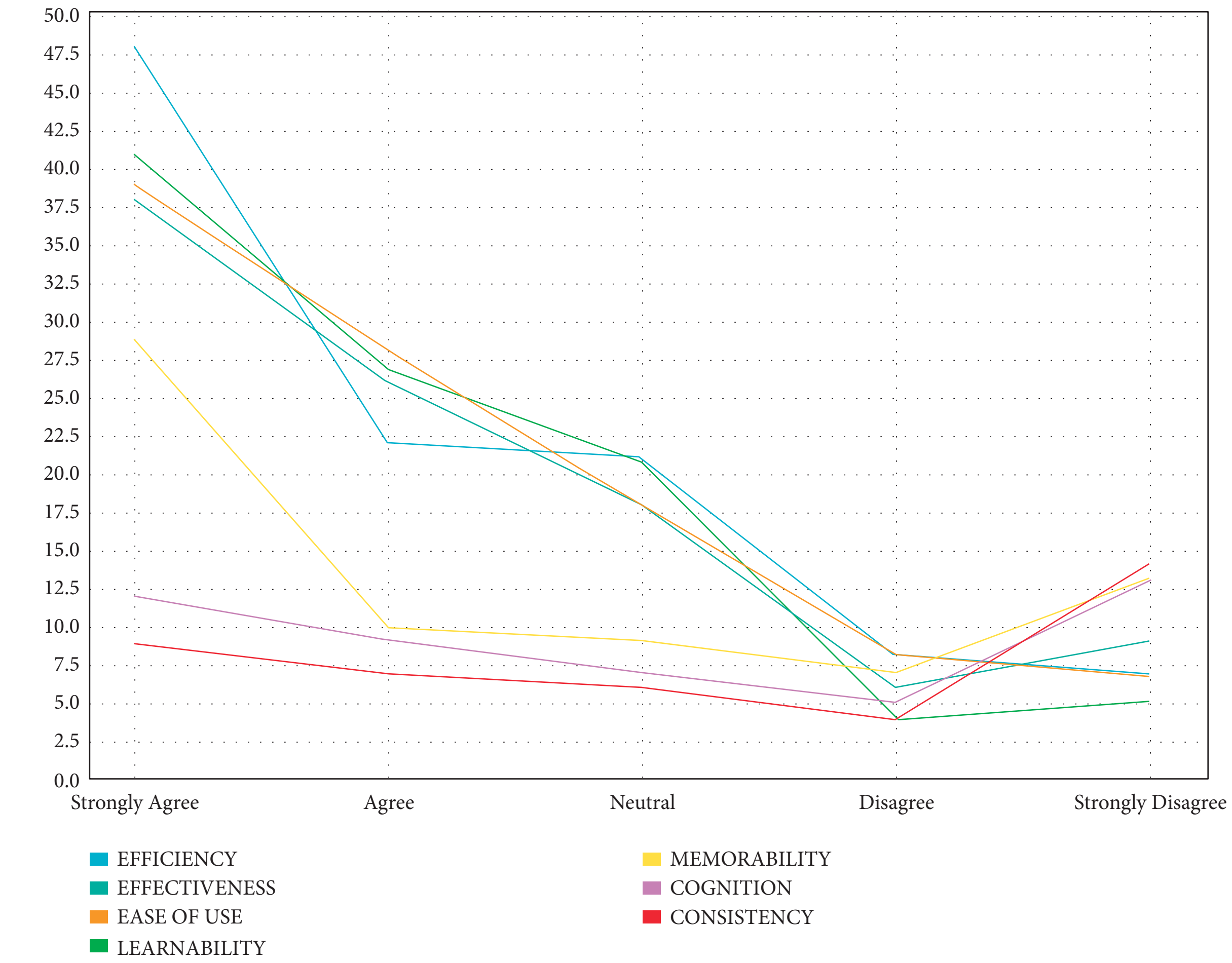

Figure 5: All features vs all Labels, response of students.

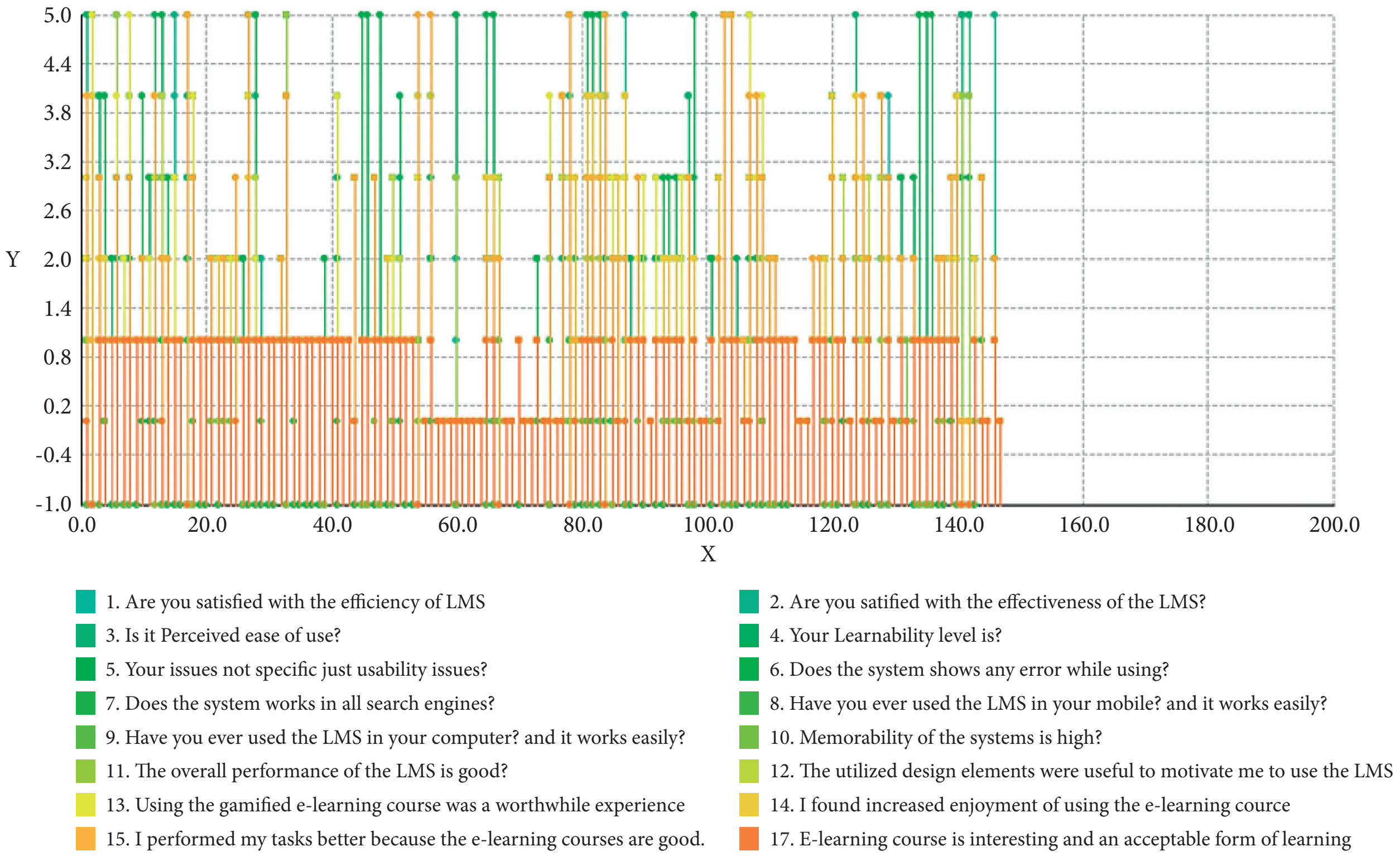

Figure 6: Each question's detail for number of respondents is presented.

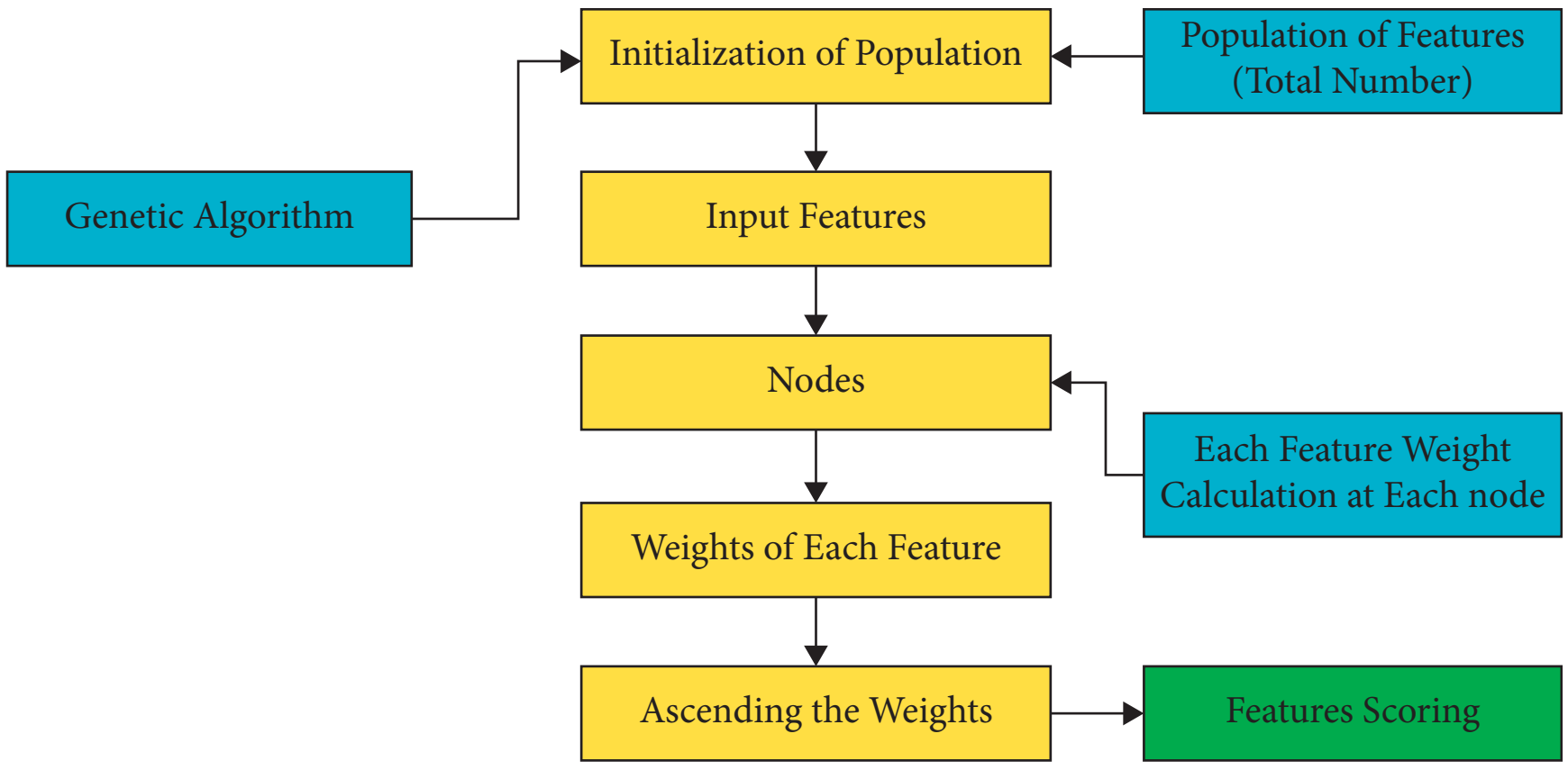

FIGURE 7: Scoring of features using GA [24].

where $nx$ is the number of unknown nodes along the $x$ -direction and $my$ is the number of unknown nodes along the $y$-direction.

*3.3.5. Feature Scoring on Ascending Order of Weights.* Then, an appropriate fitness function can be given as [7]

$$F = 11 + R. \tag{4}$$

This means that the larger the residues value the smaller the fitness one. For the exact solution $R \approx 0$ and $F \approx 1$, as the final value of F defines the score of each feature.

*3.4. Feature Clustering by Genetic Algorithm-Based Support Vector Machine.* As we have used supervised and unsupervised classifiers for our data validation, so in this step we train our unsupervised learning classifier with unlabelled data. And send the results to the normalized database for clustering. First, we have used the genetic algorithm in this approach and then applied the support vector machine classification model to predict the best features.

*3.4.1. Multiobjective Genetic Algorithm.* Multitarget GA is a population-based solution. It is useful for addressing multitarget challenges and interface difficulties. A GA may be adapted to find various solutions that are not affected by each other in one round. The capacity of GA to scan simultaneously multiple areas of solution space and can identify numerous solutions for difficult multimodal, discontinuous and nonconvex solutions. GA crossover operator may accomplish decent solutions systems with diverse goals to create new solutions that are not affected by unknown areas of the Pareto front. In comparison, the operator would not use any of the multiobjective GA to scale and prioritize goals. GA was thus the most popular heuristic approach used in our work for optimizing features.

*3.4.2. Support Vector Machine.* In SVM classification, original input values are calculated into advanced dimensional features in which classifier gives the result of the type of features. Due to these properties, SVM classifiers are inclined to possess a dressed aptitude for detecting the type of usability feature. An SVM classifier gives good results for the type of feature because some parameters in the classifier are lengthy calculated, critical, and timely taking task in the classification. Classification is done by starting with the additional discriminating features and slowly adding fewer discriminating features. Features for classification of the usability features are homogeneous, contrast, correlation, mean, and probability. SVM belongs to the supervised classification class. The leading benefits of SVM are calculated controllability, high precision, and direct decision on geometry.

Traditional SVM uses the hyperplane to classify data. In the SVM kernel, the procedure is almost similar, but the nonlinear kernel function replaces every point created between the vectors. Figure 8 shows a generic approach of an SVM model.

The gamma constraint describes the effect of a particular training sample spread with low standards. The 'C' parameter trades of misclassification of training samples in contradiction of the simplicity of the assessment surface. In SVM with no kernels, the grid examination delivered by 'GridSearchCV' thoroughly produces candidates from a grid of constraint values quantified with the tuned parameter. When 'fitting' it on a dataset all the combinations of constraint values are assessed, and the finest combination is recollected. Randomized CV implements a randomized search done by parameters, where each setting is examples from a distribution over likely parameter value.

Figure 9 shows the generic view of Machine Learning SVM Model. The validation score is shown in Figure 9; it will be transferred for interpretation or visualization. Basic terminology of our proposed genetic algorithm is shown as follows.

Optimization involves identifying input values to achieve the "bes" output values. The concept of "best" differs from problem to problem, but it implies that one or more objective functions can be maximized or reduced by changing the input parameters. The search field is made up of all potential alternatives or values that the inputs will take. There is a point or collection of points in this search region which gives the ultimate solution. The goal of optimization is to locate the search space for that point or group of points.

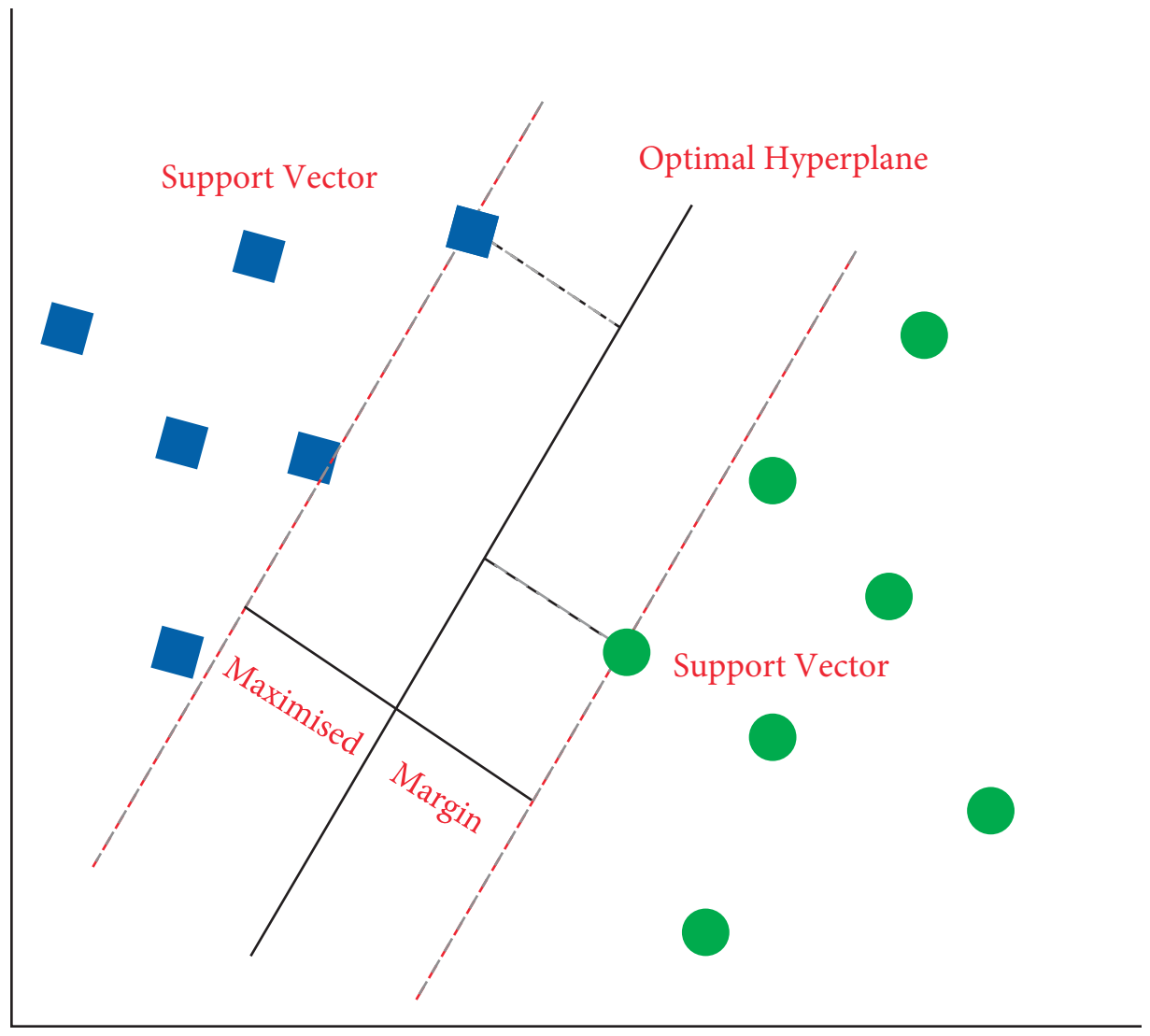

Figure 8: Generic approach of machine learning SVM model.

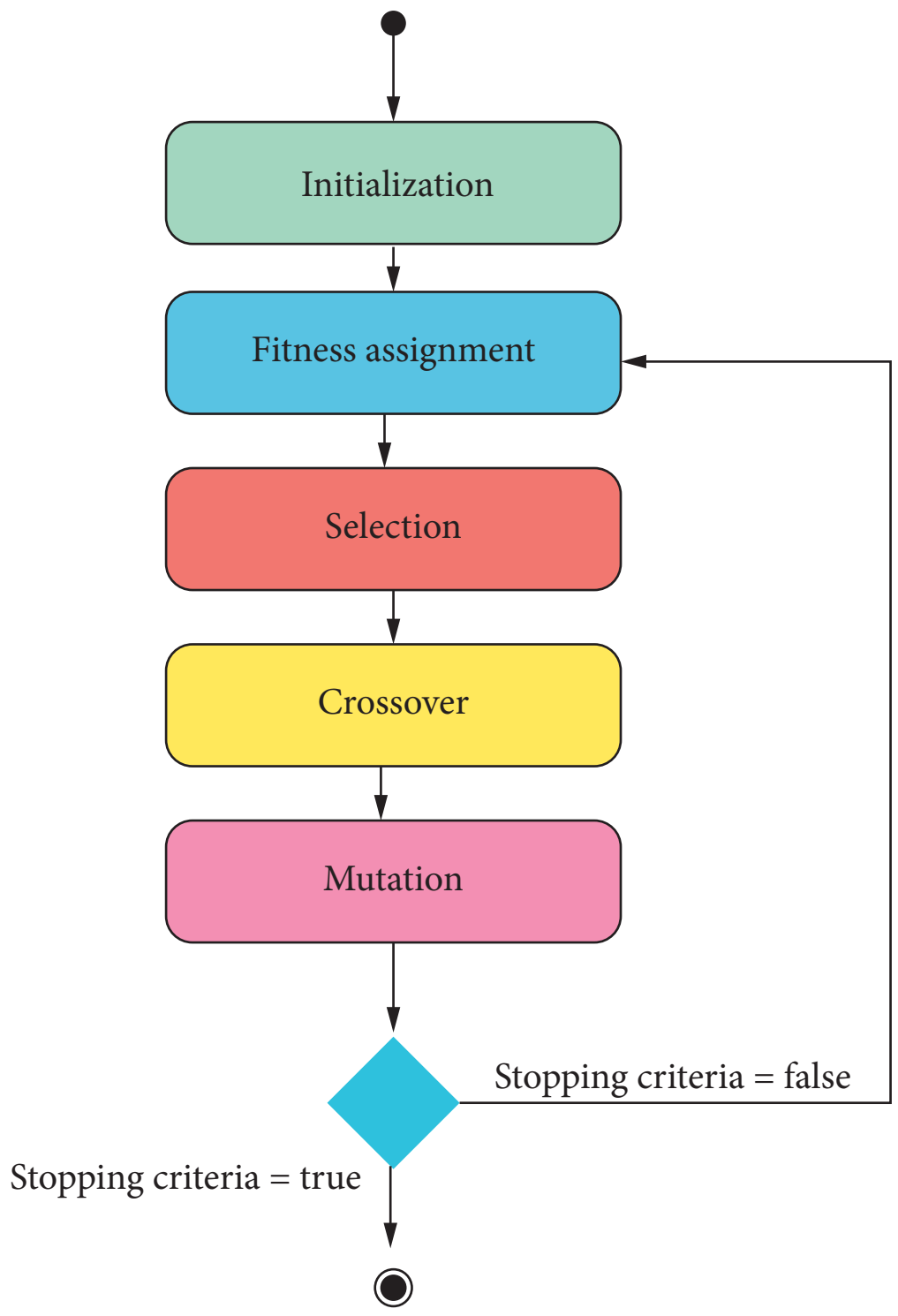

Figure 9: Basic terminology of our proposed genetic algorithm.

Nature has always been a wonderful inspirational inspiration for all humankind. Genetic algorithms (GAs) are search-oriented algorithms based on normal and genetic selection principles. GA is a subset of a far broader computing branch known as Evolutionary Computing. John Holland, students and colleagues at the University of Michigan and David E. Goldberg created GAs, which has since been tried with a strong degree of success on numerous optimization concerns. The following is the pseudo-code for GA to calculate and extract characteristics.

Figure 10 shows the basic view of GA-based support vector machine model, in which we take two split datasets, training data and forecasting data, training initiate with some value. After it we will calculate the fitness value and then create cross over function to show variation in the data, Then from stopping rule we check the parameters of parameters are complete then optimize the parameters otherwise put it into again calculate the fitness function. Otherwise predict the usability features with GA-based SVM.

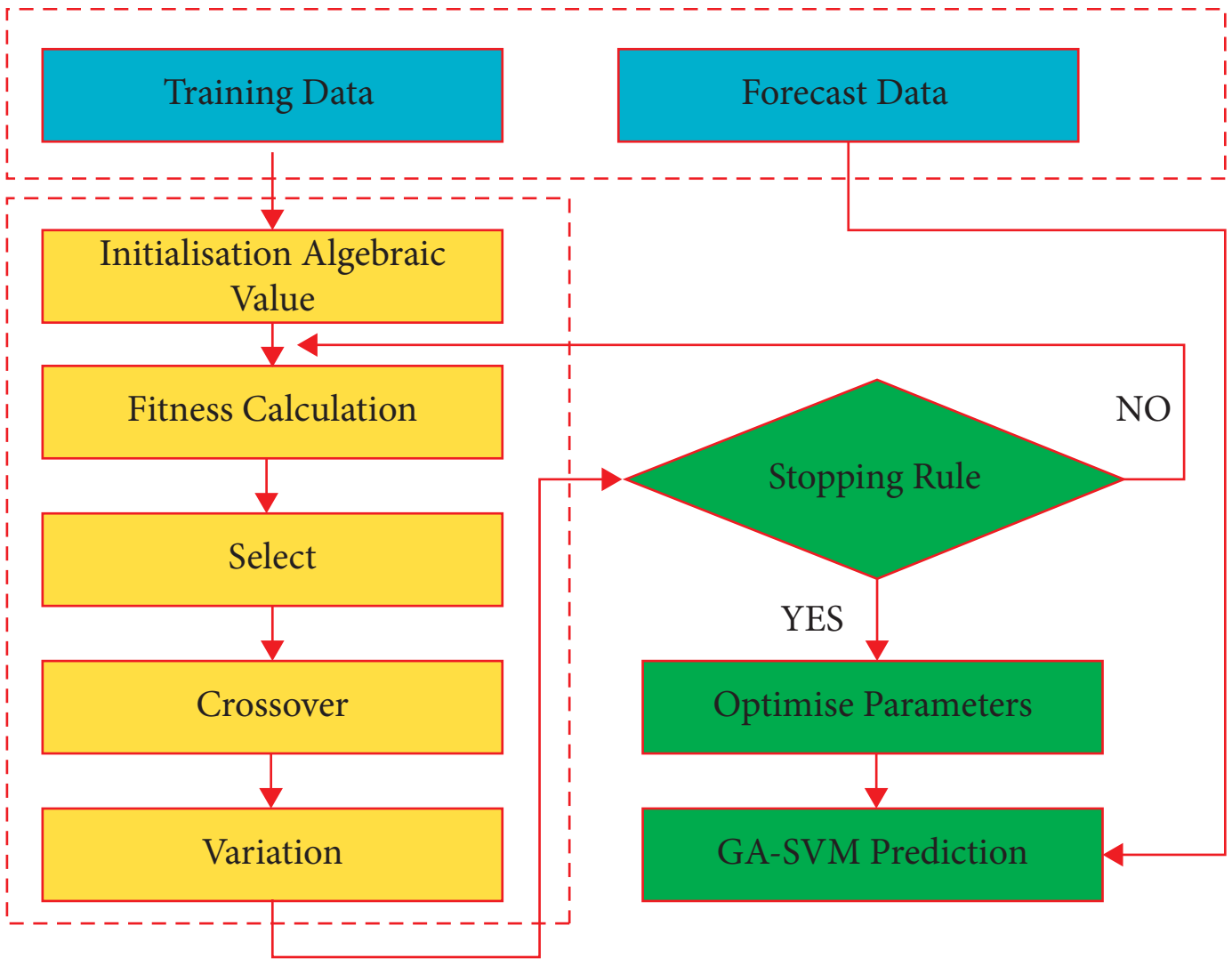

FIGURE 10: GA-based support vector machine model.

*3.5. Ranking Quality of Identified Features.* The GA is an unlabelled learning algorithm for the collection of functions. It is a network that is optimized. Consider the D-dimensional $X = \{x_1, x_2 \ldots, XD\}$ data collection, where $D$ is the sum of variables presented to the input layer. The GA is attempting to restore $X$ on the output layer. This implies that the identity function $f(x) = x$ is modelled. For this purpose, a compressed, weighted representation of the data $X$ displayed in the input layer must be retrieved from the hidden layer and then reconstructed as $X\hat{X}$ on the output layer. The GA is perfect for tasks such as reducing dimensionality and collection of features since it provides this compressed data representation.

The learning mechanism relies on the GA design for compact representation. The optimum design results in the smallest reconstruction error for both parameters (RMSE). The understanding of the weights combining on the secret layer is not inherently easy. We have taken a clear approach in our case. The weight of a vector $d$ at the node $j$ showed its value when this node was triggered. The larger a variable's weight, the more necessary it was to trigger. Thus we found the average vector weight in all the $J$ nodes, $\hat{w}$dj = totalJJj = 1wdj$\hat{w}$dj = totalj = 1Jwdj. A low $\hat{w}$dj$\hat{w}$dj variable will be less relevant than a higher $\hat{w}$dj$\hat{w}$dj variable. A selection threshold can be specified to consider selected characteristics of variables with weights above the threshold.

In the genetic algorithm, the following are the steps:

Step 1: determine the chromosome size, the rate of generation and mutation and the crossover value

[1] = [a; b; c; d] = [12; 05; 23; 08] = ;

Step 2: generates chromosome-chromosome population numbers and chromosome-chromosome genes with a random value initialization value

F obj [1] = paragraphs ((12 + 2 ∗ 05 + 3 ∗ 23 + 4 ∗ 08) − 30) = subparts ((12 + 10 + 69 + 32) − 30) = subparagraphs (123–30) = 93 [1].

Step 3: phase measures 4-7 before the generations are fulfilled

[1] = 1/(1 + F obj [1]) = 1/94 = 0.0106 = 1/0.0106.

Step 4: assessment of chromosome health by measuring objective function

Chromosome [1] = [02; 05; 17; 01] chromosome.

Step 5: selection of chromosomes

Total gen = number of gen in chromosomes ∗ population number = 4 ∗ 6 = 24.

Step 6: overlap

Chromosome [1] = [02; 05; 17; 01] chromosome.

Step 7: shift

Total gen = number of gen in chromosome ∗ population number = 4 ∗ 6 = 24.

Step 8: solution (best chromosomes)

A +2 $b$ + 3 $c$ + 4d = 30 7 + (2 ∗ 5) + (3 ∗ 3) + (4 ∗ 1) = 30 we can see that genetic algorithm variables a, $b$, c and $d$ can be equal.

We have a reservoir or community of potential alternatives to the specific issue of Coal. These solutions are also recombined and mutated (like in natural genetics), making fresh offspring and the method persists for many centuries. Each entity (or candidate solution) has a fitness value assigned (based on its objective operational value) and the fitter person has a better probability of matting and creating fitter individuals. Figure 11 shows each feature description with its score value line graph. While Figure 12 shows the feature vs score. It is aligned with the Darwinian hypothesis of "the Fittest's survival." The following is the set of features picked in Table 6 for the GA:

## 4. Experimental Setup

In the experiment phase, the collected data in the raw form is extracted from the questionnaires regarding mobile usage experiences, observations, and tests. To experiment with the used approach, the data is converted into a data-frame for processing. The labels and features are added to the data to process it. Further features like word count and weights, etc.,

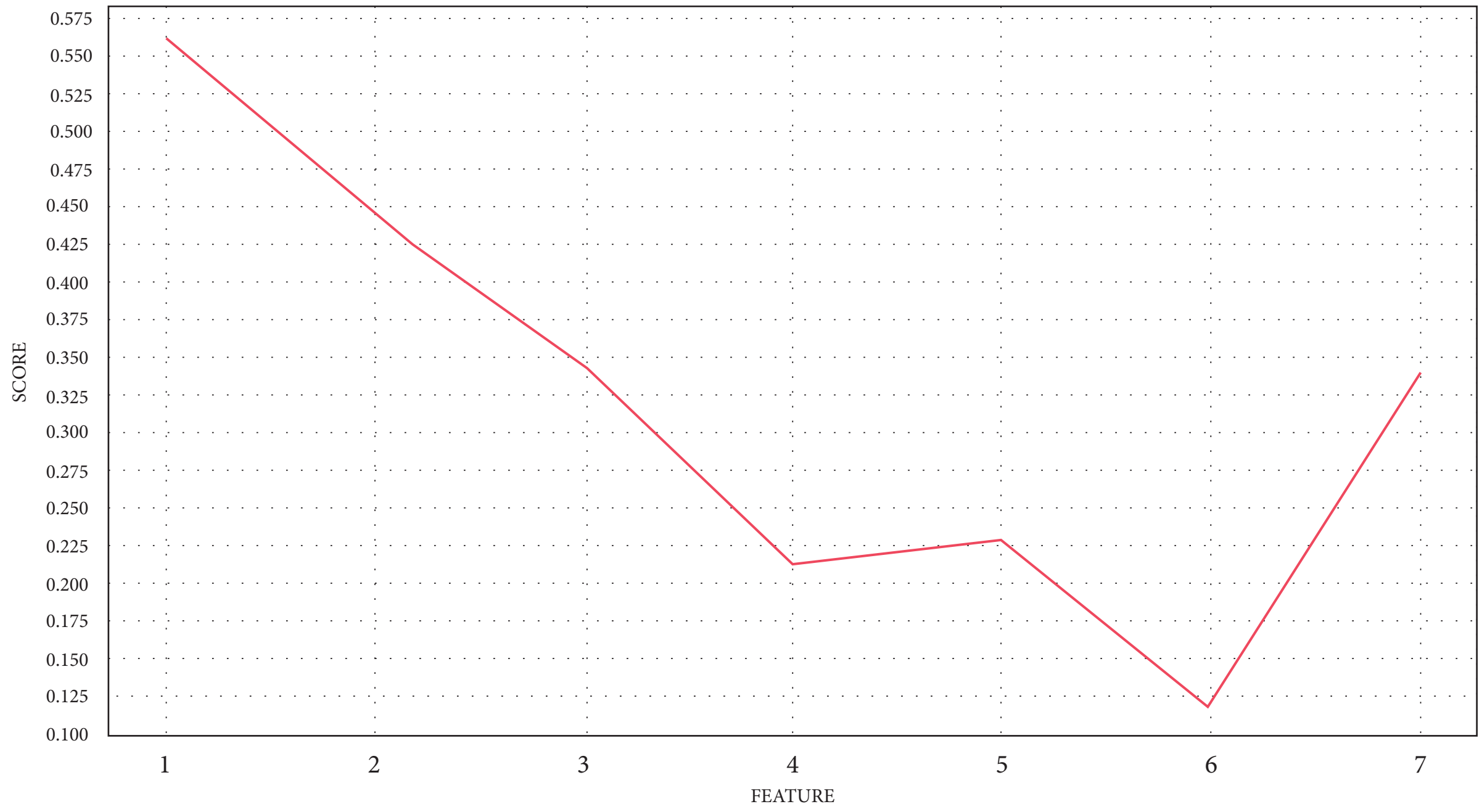

Figure 11: Each feature description with its score value line graph.

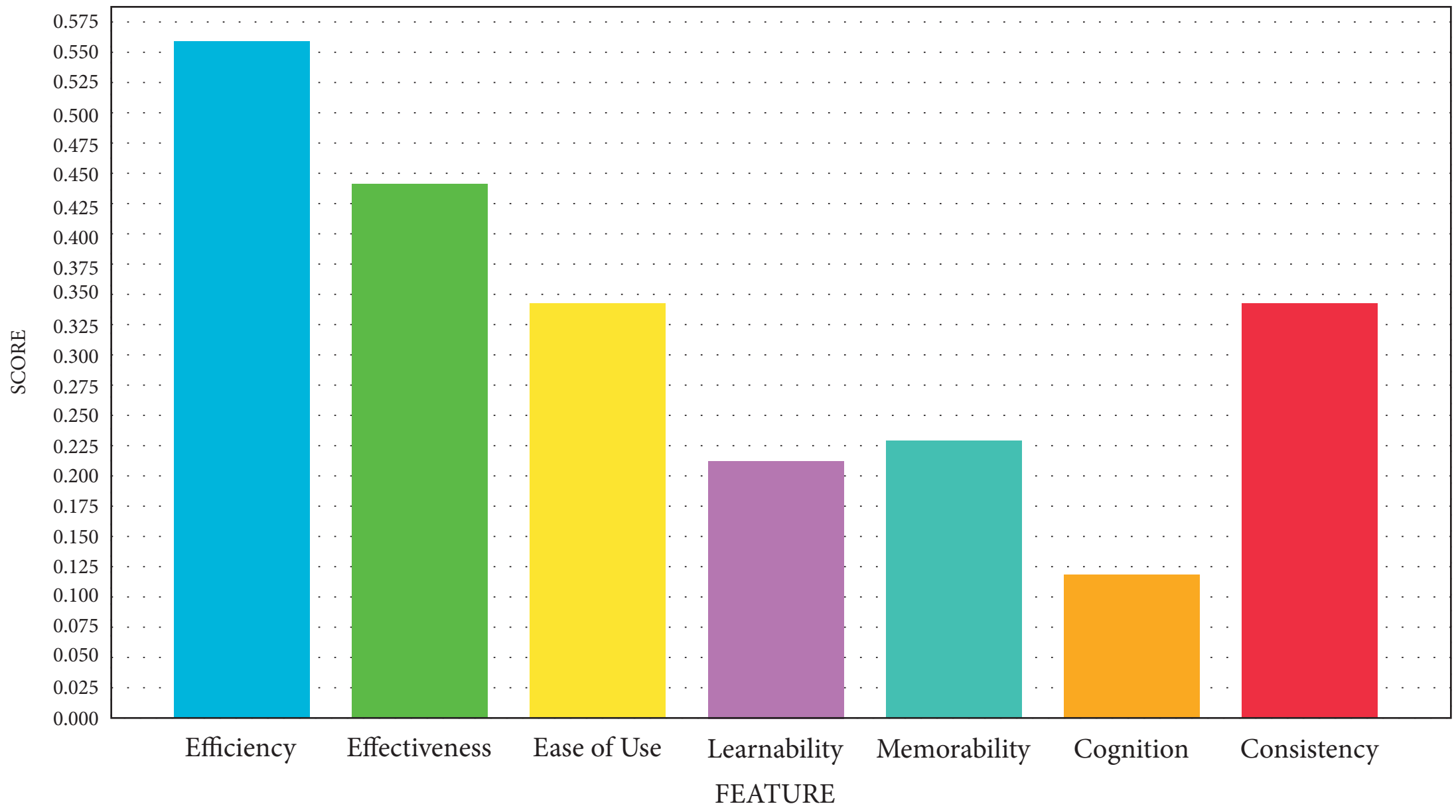

Figure 12: Score vs features bar graph.

Table 6: Scoring of features by GA.

| Feature | Score |
|---|---|
| Efficiency | 0.56 |
| Effectiveness | 0.4435 |
| Ease of use | 0.343 |
| Learnability | 0.2134 |
| Memorability | 0.23 |
| Cognition | 0.12 |
| Consistency | 0.34 |

are also added to improve the quality of data. These features are added in a CSV data file.

*4.1. Manual Labelling.* The selected dataset is used for manual labelling, and each data frame or feature is labelled with a score. Scores of each label are combined. All combined scores are converted to categories. Furthermore, these categories are added to tweak classifier parameters.

Table 7: Analysis of usability attributes of LMS-IUB (http//www.iub.edu.pk) applications.

| Features | Scoring | Benchmark | GA selection | Classification (%) | Usability analysis |
|---|---|---|---|---|---|
| Efficiency | 8 | 8.5 | −0.5 | 93 | Good |
| Effectiveness | 8 | 7 | +1 | 92 | Very good |
| Ease of use | 7 | 8 | +0.3 | 89 | Average |
| Learnability | 6 | 6.6 | +0.2 | 95 | Very good |
| Memorability | 5 | 7 | −0.2 | 93 | Fair |
| Cognition | 5 | 6 | −0.1 | 92 | Good |
| Consistency | 4 | 5 | +2 | 92 | Very good |

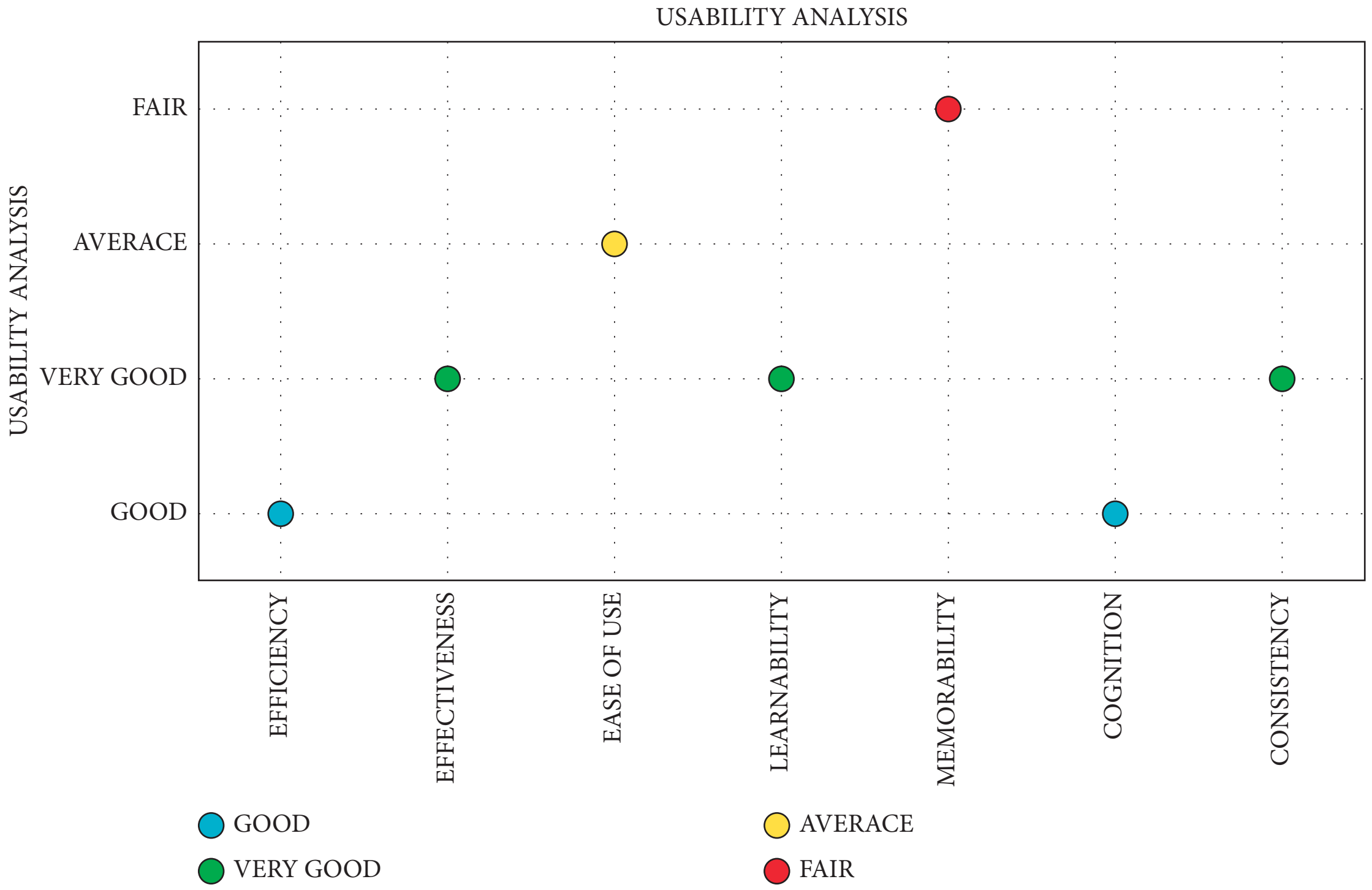

Figure 13: Ranking of usability features in terms of very good, good, fare, and average.

*4.2. Classifier Training.* In this step, the dataset is split into two parts. The first part is a training set and the second is the testing set. After this, we create an object. All classifier parameters are placed in this object. Categories of all labelled data also added to tweak classifier parameters. We fit our model for each object. After training it on the training set, we run our model (classifiers) on the best data set. We will validate the performance of our classifier based on some parameters. If the classifier's efficiency is more than 80%, then we will run our classifier on unlabelled data, otherwise, train it again on labelled data.

*4.3. Machine Learning Models.* Machine learning is used to predict and classify different datasets. Models of machine learning are parameterized to hypertune parameters to improve the model's predictive power [25, 26]. There are two type of Hyper tuning of parameters, i.e., Grid Search CV and Random Search CV [22, 27].

*4.4. Performance Parameters.* Performance parameters such as accuracy, sensitivity, specificity, and AUC are calculated for authentication of the proposed technique. The visual and parametric outcomes of the proposed technique are compared with the existing literature.

## 5. Results and Discussion

The excel sheets are compiled with GA-based SVM's python implementation to perform machine learning classification and clustering. In the first step, we have done some feature selection by using genetic algorithm. Genetic algorithms are sufficiently randomized, but they perform much better than random local search (in which we just try various random solutions, keeping track of the best so far), as they exploit historical information as well. Afterwards, machine learning models are applied for classification and clustering, to predict the usability attributes. The results are shown in Table 7.

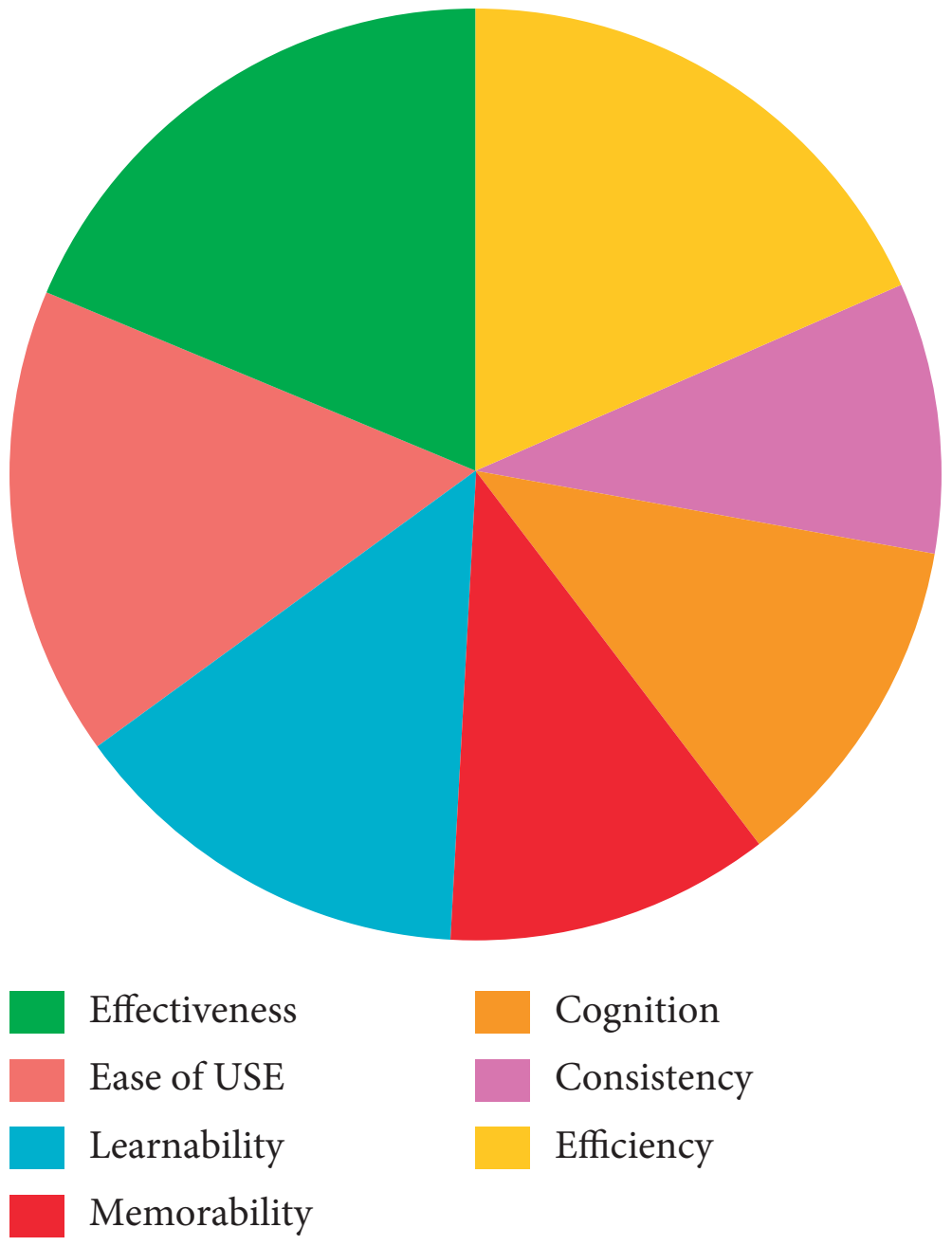

Figure 14: Pie chart showing the importance of each attribute.

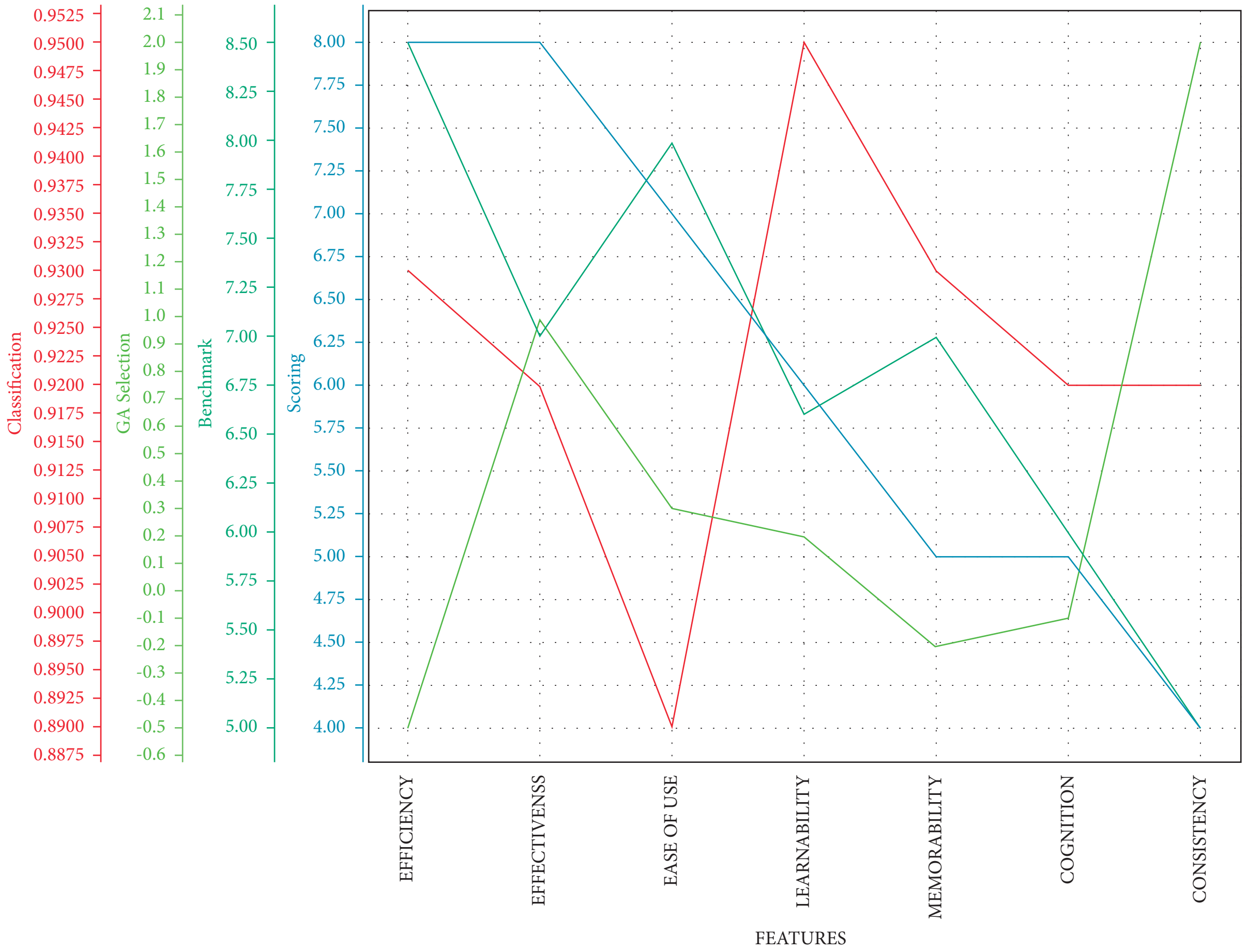

Figure 15: A variation in choices of respondents of the survey.

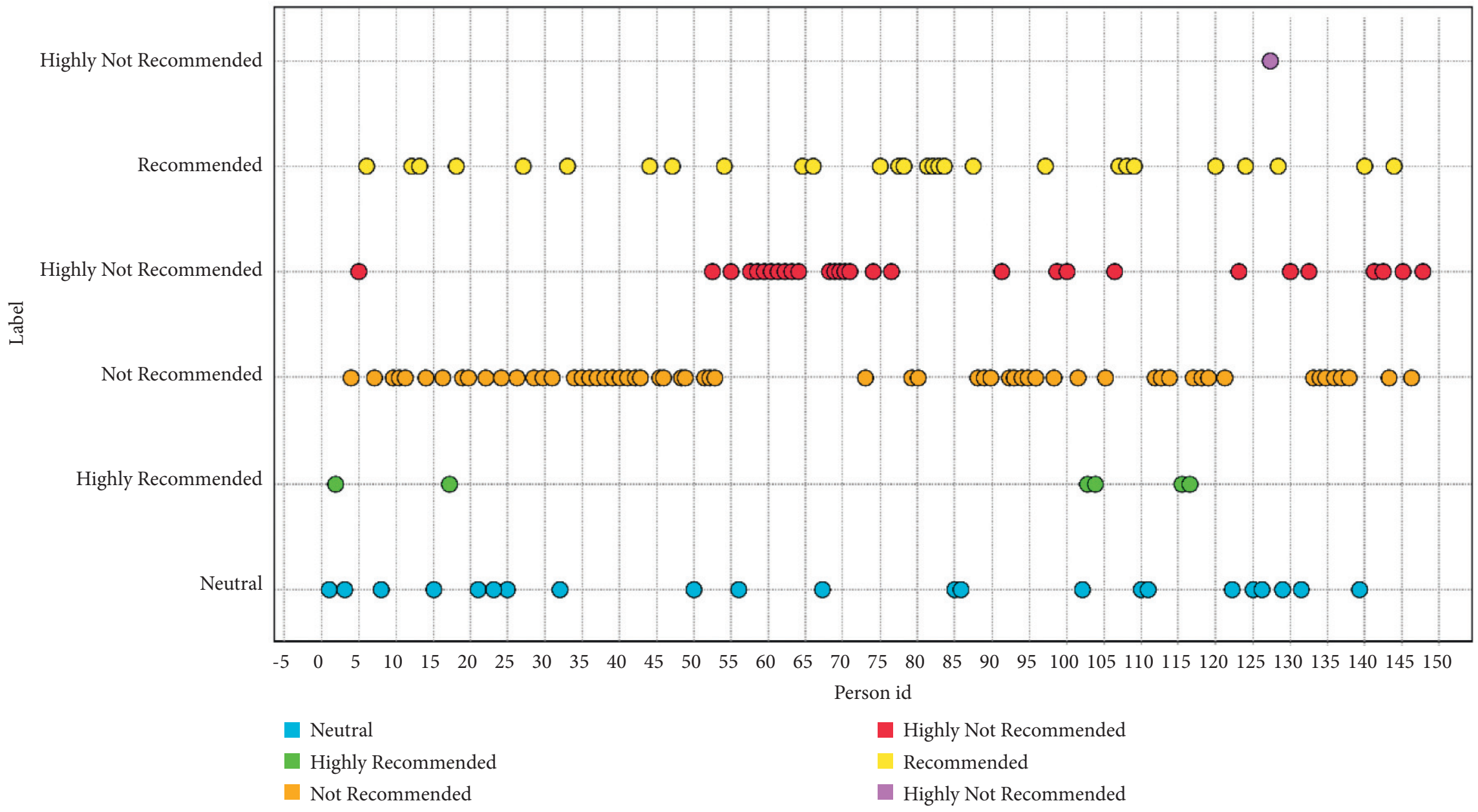

Figure 16: Each person's response is displayed in this graph for each category.

accuracy: 97.28%

| | true Neutral | true Highly Recom... | true Not Recomme... | true Highly Not Rec... | true Recommended | true highly Not Rec... | class precision |
|---|---|---|---|---|---|---|---|
| pred. Neutral | 22 | 0 | 2 | 0 | 0 | 0 | 91.67% |
| pred. Highly Recom... | 0 | 6 | 0 | 0 | 0 | 0 | 100.00% |
| pred. Not Recomme... | 0 | 0 | 60 | 2 | 0 | 0 | 96.77% |
| pred.Highly Not Rec... | 0 | 0 | 0 | 26 | 0 | 0 | 100.00% |
| pred. Recommended | 0 | 0 | 0 | 0 | 28 | 0 | 100.00% |
| pred.highly Not Rec... | 0 | 0 | 0 | 0 | 0 | 1 | 100.00% |
| class recall | 100.00% | 100.00% | 96.77% | 92.86% | 100.00% | 100.00% | |

Figure 17: Accuracy of used model.

Table 7 also shows the analysis of usability attribute of LMS-IUB application concerning features, scoring of features, their benchmark with previous studies, and feature selection by GA and classification accuracy of machine leaning models. In the last column analysis of usability by the recommendation of user has been shown as well. Figure 12 shows the visual representation of the achieved results. The graph is highlighting that feature like effectiveness learnability and consistency have good score in LMS-IUB m-learning application. In contrast, memorability feature has fair quality while ease of use is at average. Such classification will help the designers to improve various aspects of m-learning applications. Since, the users of LMA-IUB are more interested and keener in features like efficiency and ease of use, the design and suability issue related to these aspects should be improved for the better experience of the users.

Figure 13 shows the predictive analysis of the machine learning models, shows that the features which students very highly recommend are Efficiency, effectiveness and ease of use. The features like learnability and memorability come at secondary choice. Further analysis was made using GA-based SVM to find that these selected features have what quality in the targeted m-learning application.

Figure 14 shows the classification results along the GA-based selection of key features. The benchmark data is also shown in Figure 14. The graphical representation shows that bench mark data focuses on efficiency, ease of use, and memorability features of a good m-learning application in terms of better usability. However, the results of our GA-based SVM approach suggest that effectiveness and learnability are almost in line; however, the LMA-IUB m-learning application needs to improve its efficiency, ease of use, and memorability in a major way.

In Figure 15, features concerning classification, GA selection, and scoring values are plotted in series graph.

Figure 16 shows the recall and precision of the GA-based SVM approach's results for clustering dominant features in data and ranking of these dominant features. Overall, accuracy of the used approach is more than 90%.

Figure 17 shows the diversity of the data collected in survey of various subjects. It is shown in the graph that majority of the students that are user of LMS-IUB are facing issues with the efficiency, ease of use, and memorability of the m-learning application.

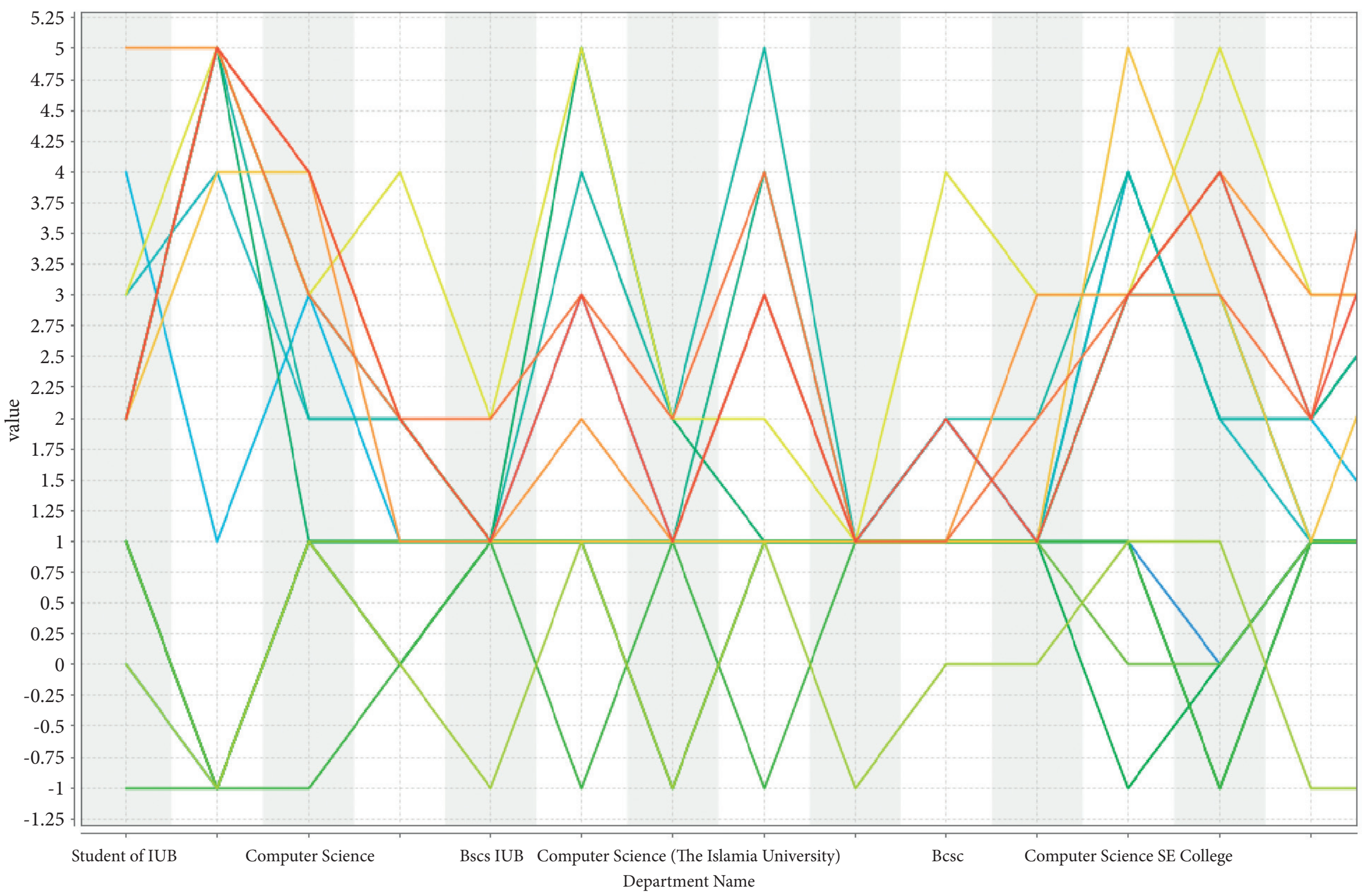

FIGURE 18: Series graph explains each question concerning department.

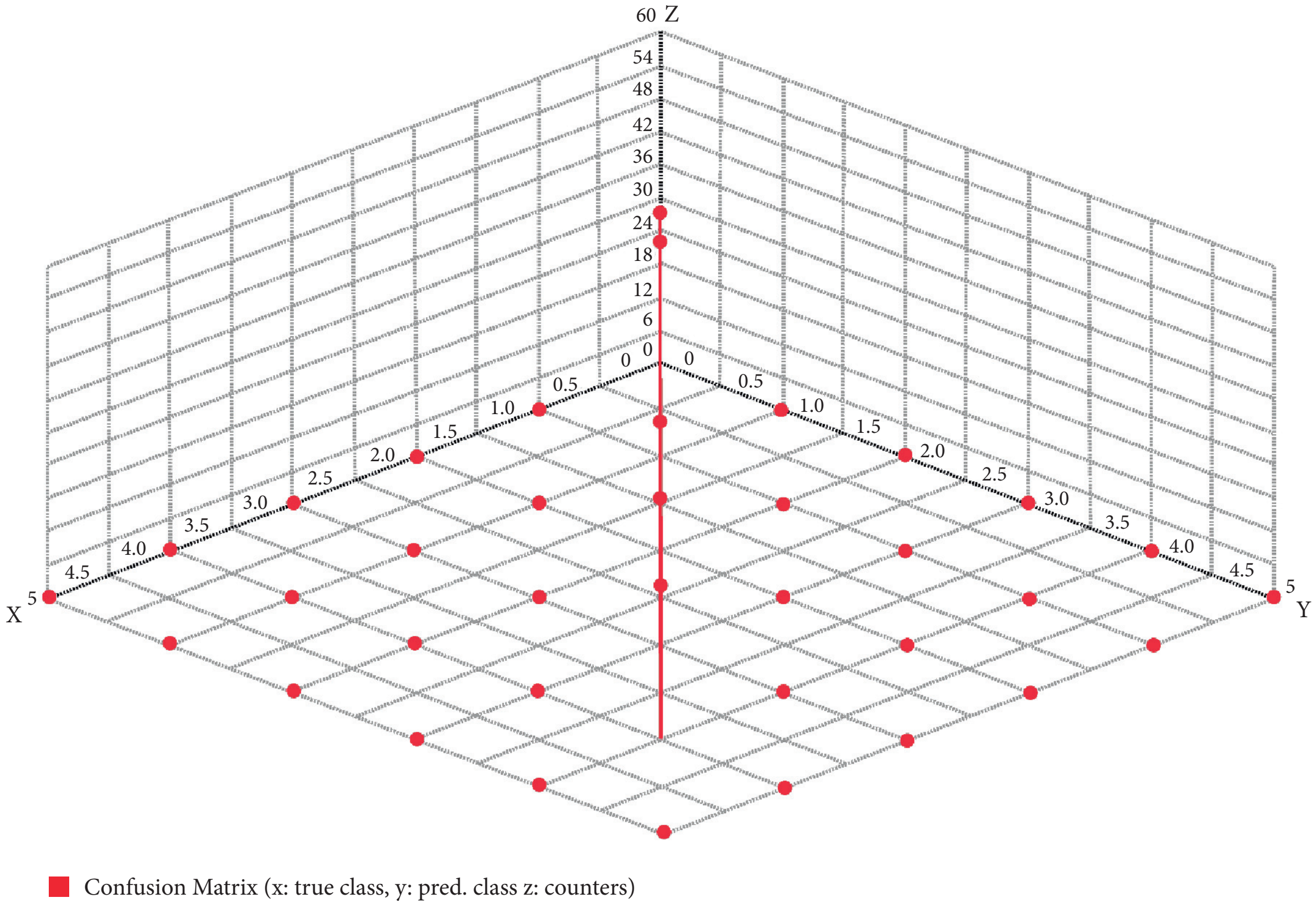

FIGURE 19: Plot view of system performance matrix.

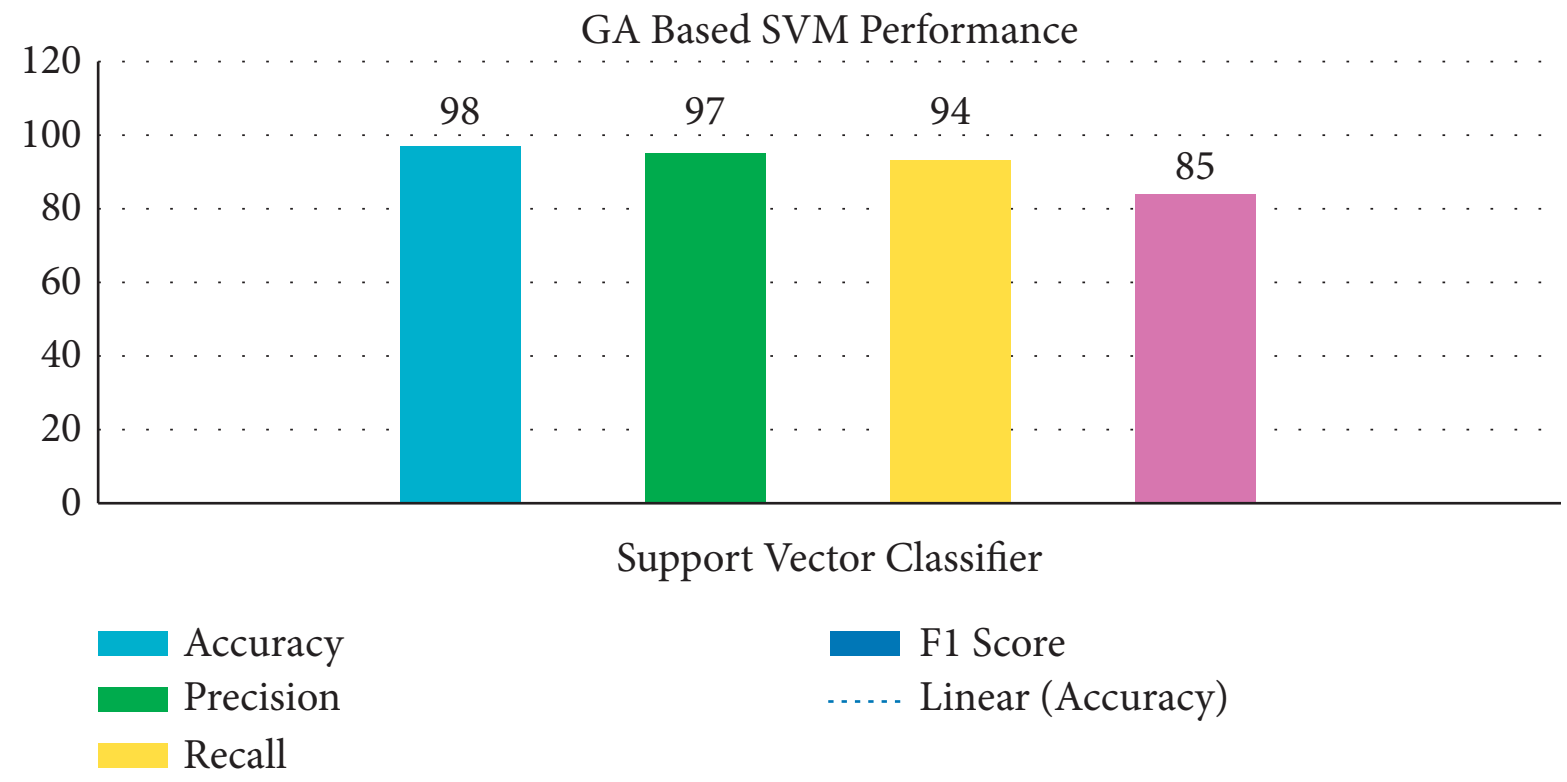

Figure 20: Performance of support vector machine.

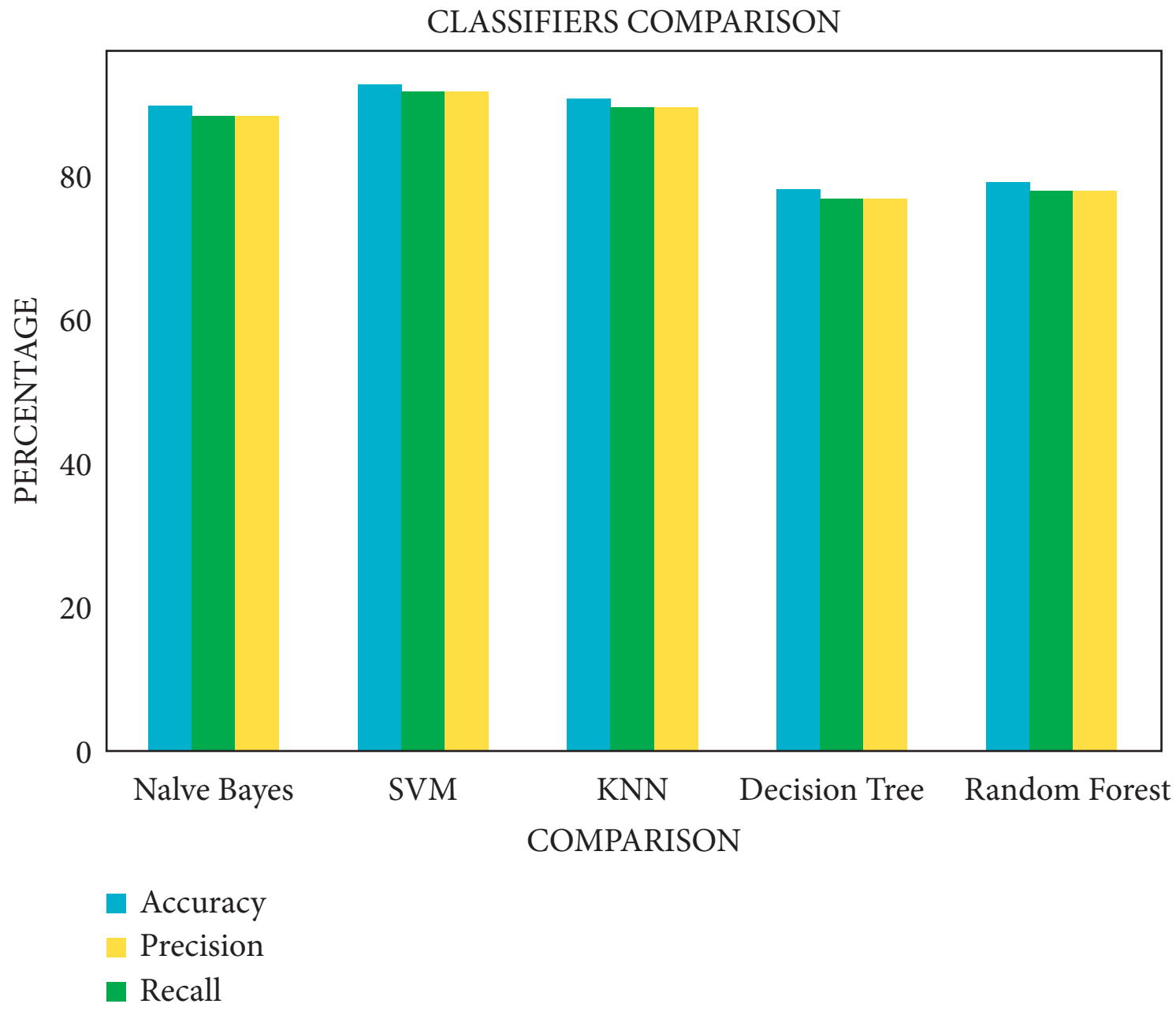

Figure 21: Comparison between different classifiers.

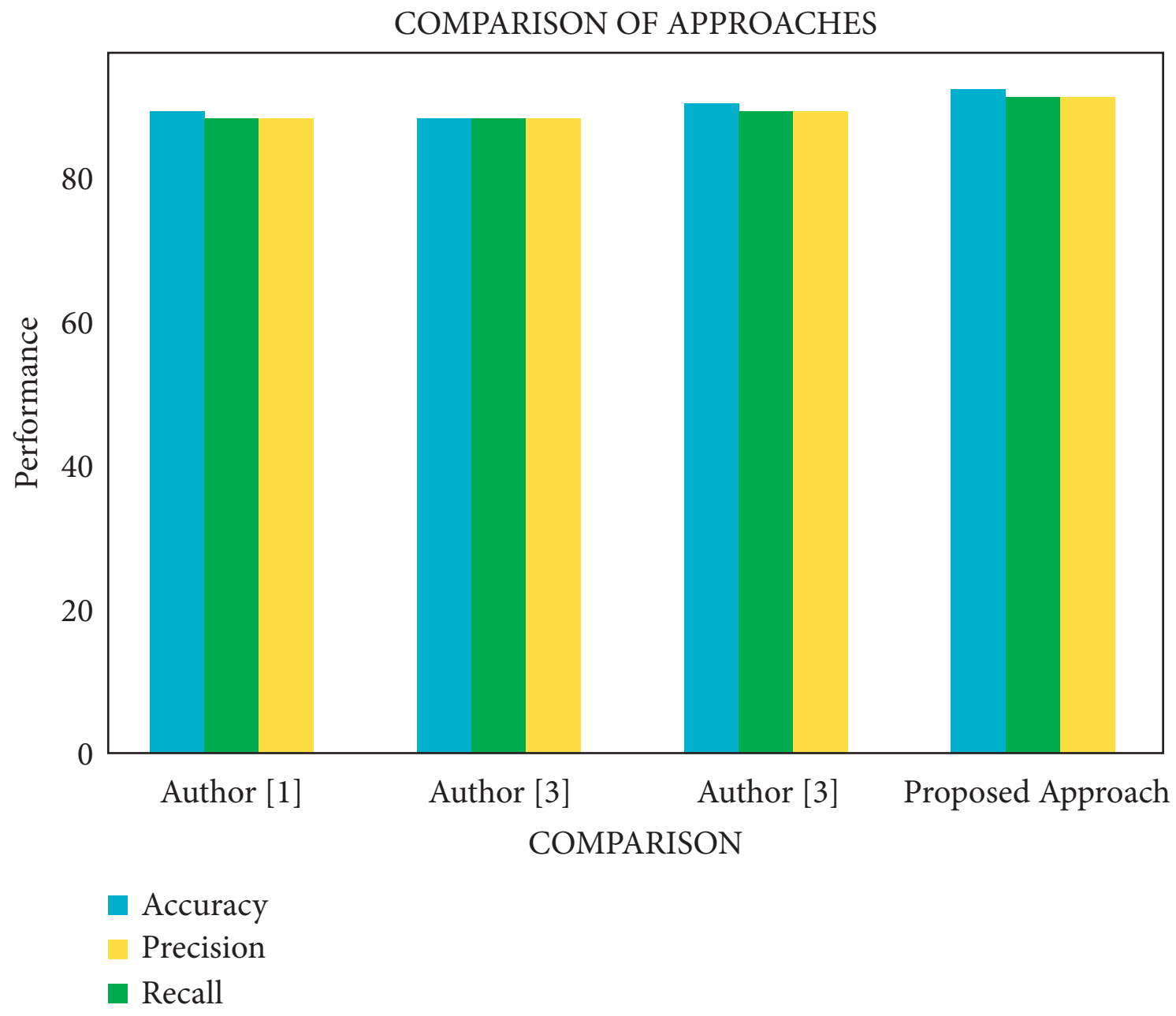

Figure 22: Comparison with Different models.

Figure 18 shows a performance matric of the proposed approach. The scatternets in the user's choice is highlighted in the figure.

Figure 19 shows the confusion matrix plot of our proposed model where true positive rate and false positive rate has been shown.

Figure 20 shows the performance of machine learning model of GA-based support vector machine to predict the usability features with 98% accuracy. Following are the prediction results of six most common machine learning models at the learning rate of 70%. Figure 21 shows the comparison between different classifiers.

Following Figure 22 shows the comparison of the results of the various models with the used model of GA-based SVM. It is highlighted in Figures 21 and 22 that the proposed GA-based SVM performs better than the other approaches.

The best among these is support vector machine with the prediction score of more than 90% which is consistent in other parameters. We have also compared our work with the previous models regarding accuracy, as shown in Figure 22.

## 6. Conclusion

This paper focuses on identifying quality issues with usability of m-learning applications being used in Pakistani educational institutes. The purpose is to facilitate the adaptation of digital technologies in education sector more easily. This work performs the quantitative and qualitative analysis of the usability features of m-learning applications. In this study, a questionnaire-based survey is conducted to find out the mental perception towards GUI of an m-learning application. This work identifies the key elements in usability of m-learning applications and recommends the design heuristics that are to be adapted for the better design of such applications. Considering a user's (student) perception, it has been analysed that efficiency has been the top-ranked usability feature by which the student is satisfied. In this paper, genetic algorithm (GA) based support vector machine (SVM) is used to score the features of usability and extract dominant features. Each usability feature has been scored according to the weights of each feature of usability. The rank of main features is calculated and used in the questionnaire and analysed with respect to a benchmark using the genetic algorithm-based SVM approach. After that, GA-based SVM is used for feature selection. GA-based SVM identified the dominant features. Based on the high ranked features, further recommendation based on the usability features is made concerning students' perception. Based on the prediction of machine learning models, GA-based SVM is considered a more efficient model in predicting the best features with 90% accuracy [28].

## Data Availability

Data of this article will be available if required.

## Conflicts of Interest

The authors declare that they have no conflicts of interest.